\documentclass[preprint2]{aastex63}

\def\gsim{\mathrel{\rlap{\lower 4pt \hbox{\hskip 1pt $\sim$}}\raise 1pt
\hbox {$>$}}}
\def\lsim{\mathrel{\rlap{\lower 4pt \hbox{\hskip 1pt $\sim$}}\raise 1pt
\hbox {$<$}}}

\received{}
\revised{}
\accepted{by ApJ on 24 December, 2021}
\shorttitle{Properties of SNe Ibn}
\shortauthors{Maeda \& Moriya}
\graphicspath{{./}{figures/}}
\begin{document}

\title{Properties of Type Ibn Supernovae: Implications for \\
the Progenitor Evolution and the Origin of a Population of Rapid Transients}

\correspondingauthor{Keiichi Maeda}
\email{keiichi.maeda@kusastro.kyoto-u.ac.jp}

\author[0000-0003-2611-7269]{Keiichi Maeda}
\affiliation{Department of Astronomy, Kyoto University, Kitashirakawa-Oiwake-cho, Sakyo-ku, Kyoto, 606-8502. Japan}

\author[0000-0003-1169-1954]{Takashi J. Moriya}
\affiliation{National Astronomical Observatory of Japan, National Institutes of Natural Sciences, 2-21-1 Osawa, Mitaka, Tokyo 181-8588, Japan}
\affiliation{School of Physics and Astronomy, Faculty of Science, Monash University, Clayton, Victoria 3800, Australia}



\begin{abstract}
Type Ibn Supernovae (SNe Ibn) show signatures of strong interaction between the SN ejecta and hydrogen-poor circumstellar matter (CSM). Deriving the ejecta and CSM properties of SNe Ibn provides a great opportunity to study the final evolution of massive stars. In the present work, we present a light curve (LC) model for the ejecta-CSM interaction, taking into account the processes in which the high-energy photons originally created at the forward and reverse shocks are converted to the observed emission in the optical. The model is applied to a sample of SNe Ibn and `SN Ibn' rapidly evolving transients. We show that the characteristic post-peak behavior commonly seen in the SN Ibn LCs, where a slow decay is followed by a rapid decay, is naturally explained by the transition of the forward-shock property from cooling to adiabatic regime without introducing a change in the CSM density distribution. The (commonly-found) slope in the rapid decay phase indicates a steep CSM density gradient ($\rho_{\rm CSM} \propto r^{-3}$), inferring a rapid increase in the mass-loss rate toward the SN as a generic properties of the SN Ibn progenitors. From the derived ejecta and CSM properties, we argue that massive Wolf-Rayet stars with the initial mass of $\gsim 18 M_\odot$ can be a potential class of  the progenitors. The present work also indicates existence of currently missing population of UV-bright rapid transients for which the final mass-loss rate is lower than the optical SNe Ibn, which can be efficiently probed by future UV missions.  
\end{abstract}



\section{Introduction} \label{sec:intro}

Core-collapse supernovae (CCSNe) are an explosion of a massive star with the zero-age main-sequence mass ($M_{\rm ZAMS}$) of $\gsim 8 M_\odot$ at the end of their evolution, following the exhaustion of the nuclear fuel and the subsequent catastrophic collapse of the central core. Classically they are divided into several observational types based on the presence and/or absence of characteristic spectral lines \citep{filippenko1997}; type II SNe (SNe II) with strong H lines, type Ib SNe (SNe Ib) showing He lines rather than H lines, and type Ic SNe (SNe Ic) having neither H nor He lines. Type IIb SNe (SNe IIb) bridge between SNe II and Ib, showing initially strong H lines that are later replaced by He lines. The sequence of SNe II-IIb-Ib-Ic is interpreted as the outcome of an increasing amount of the envelope stripping during their pre-SN evolution, which is driven either by a strong stellar wind or binary interaction \citep{langer2012}. Unlike red-supergiant (RSG) progenitors for SNe II \citep{smartt2009}, the progenitors of SNe IIb/Ib/Ic are thus believed to be massive stars that have lost at least most of the hydrogen envelope (i.e., an analog of a Wolf-Rayet star; WR); they are collectively called stripped-envelope SNe (SESNe). 

Supernovae (SNe) provide an irreplaceable opportunity to deepen our understanding on still unclarified stellar evolution in their final phases. With rapidly increasing capability of transient surveys and follow-up observations, various new insights have been obtained, some of which indeed raise a challenge to our standard knowledge on the stellar evolution and SN explosion. One of the highlights in such recent development is high-density circumstellar matter (CSM) located just in the vicinity of some (or most of) SN progenitors ($\lsim 10^{15}$ cm). This `confined CSM' was first indicated for SN IIn 1998S \citep{chugai2001}, and systematic investigation has become possible for various types of CCSNe in the modern survey and follow-up era; 
the confined CSM manifests itself in recombination emission lines of highly-ionized ions in their spectra within a day or at most a few days since the explosion \citep[`flash spectroscopy': ][]{gal-yam2014,khazov2016,yaron2017} or in early light curve evolution within the first 10 days either in the thermal emission \citep{forster2018} or in the non-thermal emission \citep{maeda2021}. For the pre-SN mass-loss wind/ejection velocity of $\sim 10$ km s$^{-1}$ \citep[for an RSG: ][]{smith2014a,moriya2017} or $\sim 1,000$ km s$^{-1}$ \citep[for a WR: ][]{chevalier2006,crowther2007}, the CSM must have been created by the mass loss in the final 30 or 0.3 years toward the SN explosion, with the mass-loss rate reaching to at least $\sim 10^{-3} M_\odot$ yr$^{-1}$ or even larger \citep{groh2014,morozova2015,moriya2017,yaron2017,forster2018}, beyond what the standard theory of stellar evolution and stellar wind predicts. Such an increasing stellar activity toward the end of a massive star's evolution has not been considered in the classical stellar evolution theory. Identifying the mechanism leading to such a pre-SN activity \citep{quataett2012,smith2014b,fuller2017,fuller2018,ouchi2019,morozova2020}, which is under active debate, will be a key to completing our knowledge of stellar evolution to provide solid basis for many branches of astronomy. 

Another important insight obtained through the recent development of transient observations is the emerging diversity of CCSN subclassses, thanks to the rapidly increasing sample that allows investigation of rare populations. A subclass of interest in the present work is a class of interacting SNe, i.e., those showing signatures of strong interaction between the SN ejecta and dense CSM \citep[see ][for a review]{smith2017}. The most famous example is the SN IIn showing strong and narrow H emission lines, which are interpreted as an SN explosion within H-rich dense CSM. The corresponding (estimated) mass-loss rate is $\gsim 10^{-3} M_\odot$ yr$^{-1}$, sometimes exceeding $\sim 1 M_\odot$ yr$^{-1}$\citep{moriya2013,moriya2014a}. A fraction of them also show pre-SN activity directly detected in pre-SN images, reinforcing the idea that the pre-SN activity of massive stars, irrespective of its unclarified origin, creates the dense CSM \citep{pastorello2007,ofek2013,ofek2014,smith2014c,strot2021}. 

There has been increasing interest in the H-poor analog of SNe IIn, the so-called SNe Ibn as characterized by (relatively) narrow (typically $\sim 1,000$ km s$^{-1}$) He emission lines \citep{matheson2000,pastorello2007}. They form a rare population (with the volumetric rate being $\sim 1$\% of CCSNe; Section 6), and thus only recently it has become possible to construct a statistical sample \citep{hosse2017}. In another view, they are also considered to be an analog of SNe Ib but with dense CSM surrounding the progenitor He star. The sites of SNe Ibn are generally associated with star-forming environment \citep{pastorello2015c}, while one exceptional, but perhaps non-negligible, case has been found with no/little association of star-formation activity \citep{sanders2013,hosse2019}; it has thus been indicated that a main route to SNe Ibn is a CCSN of a massive star, while some fraction may come through totally different evolutionary pathway. 

Based on the circumstantial evidences summarized above, a popular idea for their origin is a core-collapse of a massive WR star \citep{pastorello2007,tominaga2008}, while binary evolution may also play a role \citep{foley2007}. However, further testing the ideas has been limited by a lack of our detailed knowledge on how their characteristic observational properties are explained. Especially, despite the general idea that their light curves (LCs) are powered by the interaction between the SN ejecta and H-poor CSM, the optical LC model for SNe Ibn based on this scenario has been largely unexplored, without which the properties of the SN ejecta and the CSM cannot be quantified; while there are some LC models for SNe Ibn with different levels of sophistication \citep{tominaga2008,pellegrino2021,dessart2021}, some key processes are still likely missing (see Section 3). This is the topic of the present work; we construct the (general) LC model for the SN-CSM interaction scenario, especially addressing how the kinetic energy dissipated by the interaction is converted into the thermal radiation mostly in the optical (and UV/NIR) as we observe.

SNe Ibn have some distinguishing characteristics in their LC properties \citep{pastorello2016,hosse2017}. They are among the most luminous SNe with the absolute peak magnitude in the range of $\sim -18.5$ to $\sim -20$ mag. The rise time is generally short, with the shortest ones peaking within at most a few days of the explosion. Indeed, some of them can be classified as `rapidly-evolving transients' (see Section 6). The rapidly-evolving transients have been an underrepresented population as the classical surveys were not able to discover such a short-time scale transients, while the increasing examples rapidly constructed by the recent high-cadence surveys and follow-up activities indicate that they indeed form a non-negligible population among explosive transients \citep{drout2014,tanaka2016,pursiainen2018,tampo2020,ho2021}. As such, understanding the origin of SNe Ibn is important to map the different types of stellar explosions to the rapidly expanding zoo of the transient sky. 

Unlike SNe IIn which tend to evolve slowly, SNe Ibn generally show a very rapid decay in their LCs after the peak. While there are some outliers \citep{pastorello2015b}, the LCs of SNe Ibn are very homogeneous and this rapid decay seems to be a common feature among SNe Ibn \citep{hosse2017}. It has been qualitatively interpreted \citep[e.g.,][]{moriya2016} that the dense CSM is confined at the vicinity of the SN progenitor (which may to some extent be a common feature of CCSNe in general; see above), but quantitative analysis is largely missing due to the lack of the reliable LC models. 
%This is one of the main topics of the present work. 

The present paper is structured as follows. To set a scene, we review the LC properties of SNe Ibn in Section 2. In Section 3, we present the LC model for the ejecta-CSM interaction, taking into account the processes in which the power dissipated at the shocks is converted to the observed radiation. General LC properties expected for such a model are clarified in terms of the hydrodynamic interaction and the radiation processes, and the dependence of the LC on the ejecta and CSM properties is investigated. In Section 4, the LC model is applied to a sample of SNe Ibn, and the properties of the ejecta and CSM are derived (or constrained). The results will then be used to infer the possible progenitor systems in Section 5, where we argue that the massive WR stars with $M_{\rm ZAMS} \gsim 18 M_\odot$ are the most promising progenitors for SNe Ibn. The model is further applied to a sample of `SN Ibn' rapidly-evolving transients in Section 6, where we predict existence of currently underrepresented UV-strong rapid transients, which are originated within the same framework but with the final mass-loss rate smaller than the presently identified SN Ibn (and rapid) population in the optical. Our findings are summarized in Section 7.

\section{Light Curve Properties of SNe Ibn}\label{sec:obs}

\begin{figure}[t]
\centering
\includegraphics[width=\columnwidth]{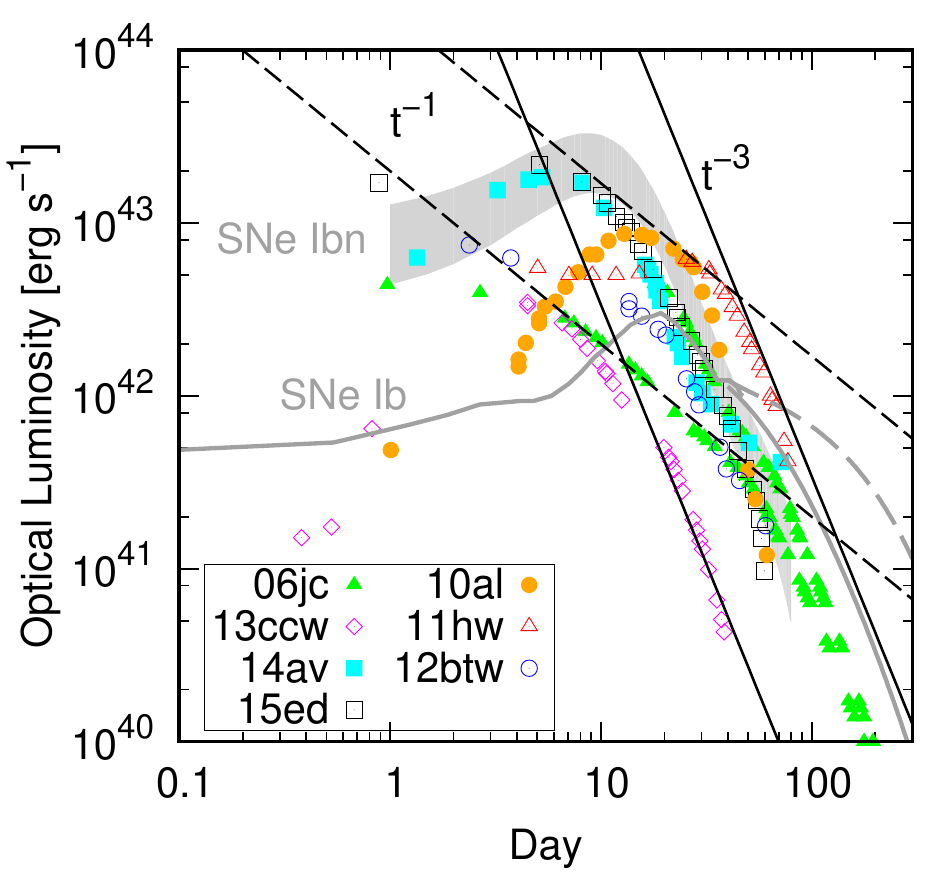}
\caption{A compilation of the (quasi-bolometric) light curves of selected SNe Ibn taken from \citet{moriya2016}. The filled symbols are those including the IR contribution while the open symbols are those including the optical contribution only. The $R$/$r$ template light curve for SNe Ibn \citep[gray area;][]{hosse2017} represents the sample behavior well. The bolometic template for SNe Ib \citep[gray-solid;][]{lyman2016} is also shown for comparison, which is then connected to the $^{56}$Co energy input with (gray-solid) or without (gray-dashed) the effect of the $\gamma$-ray optical depth decrease \citep{maeda2003}. The power-law functions are shown for $\propto t^{-3}$ (black-solid) and $\propto t^{-1}$ (black-dashed). For SN 2006jc, two cases are shown with different choice of the peak/explosion date.
}
\label{fig:lcobs}
\end{figure}

Fig. \ref{fig:lcobs} shows quasi-bolometric LCs of SNe Ibn studied in the present work, taken from \citet{moriya2016}. This is a compilation of the data from the following literature: \citet{pastorello2007,pastorello2015c,pastorello2015a,pastorello2015d,pastorello2016}. Following the quick luminosity decrease, some SNe Ibn show signatures of dust formation in the late phase; for example, a prototypical SN Ibn 2006jc started forming dust at $\sim 50$ days after the LC peak \citep{smith2008} followed by the increase in the infrared (IR) luminosity that dominates the bolometric luminosity at $\sim 80$ days after the LC peak \citep{anupama2009,sakon2009}. In Fig. \ref{fig:lcobs}, the IR contribution is taken into account for SNe 2006jc, 2010al, and 2014av (i.e., those shown by filled symbols). For the other SNe Ibn studied in the present work, the possible IR contribution is not included. However, we believe this effect is not important for the phases analyzed in the present work; for SN 2006jc, the dust formation has started once the luminosity has decreased to a few $\times 10^{41}$ erg s$^{-1}$. the other SNe Ibn in our sample keep brighter than this luminosity almost for the whole duration considered here. For example, SN 2011hw, which has a similar (optical) decay rate with the (bolometric) decay rate of SN 2006jc, did not show any signature of dust formation up to 50 days covering the whole period considered here \citep{smith2012}. Recently, \citet{gan2021} studied long-term NIR evolution of four SNe Ibn, and found three out of four form only a negligible amount of dust; one SN Ibn (OGLE-2012-SN-006) shows a signature of a large amount of dust, but this SN Ibn is an outlier in its LC evolution (showing very slow and flat evolution) \citep{pastorello2015b} and thus omitted in the present work. 

The LC behavior of SNe Ibn is rather homogeneous \citep{hosse2017} while there are some outliers \citep{karam2017,kool2021}\footnote{We have omitted OGLE-2012-SN-006 from the sample of \citet{moriya2016}, since its LC behavior is clearly an outlier and does not follow the template evolution. }. We further find that the behavior can be divided into three phases; initial rise (which is not always seen due to the observational difficulty), relatively slow decay after the LC peak, and then the rapid decay. Interestingly, the slow and fast decay phases can be described by a power-law decline in the luminosity (Fig. \ref{fig:lcobs}). In the slow decay phase after the LC peak, a typical evolution is described by $L \propto t^{-1}$. This is then followed by a rapid decay with $L \propto t^{-3}$. 
%This behavior is also seen in the SN Ibn (r/R-band) LC template \citep{hosse2017} where the post-peak slow decay is marginally seen. 

It is intriguing that the LC shape is very uniform and roughly described by a (double) power-law function, despite the diversity in the luminosity scale and time scale to the peak. The power-law behavior indicates that the underlying physical processes are described (approximately) by a combination of power-law functions, supporting the SN-CSM interaction as a main source for radiation output from SNe Ibn. However, as noted by previous studies \citep[e.g.,][]{moriya2016}, the decay slope in the late phase is much steeper than classical populations of SNe, including SNe IIn that are powered by the SN-CSM interaction in the H-rich environment. 

%To set the scene for the main modeling analysis in the present work, we start our discussion here by showing that the rapid decay raises a challenge for the standard SN-CSM interaction scenario with a CSM created by a steady-state mass loss, i.e., $\rho_{\rm csm} \propto r^{-2}$. For the standard SN-CSM interaction model under (frequently adopted) two assumptions, (1) the ejecta density structure is described by a single power law (or the reverse shock is in the outer ejecta with a steep density gradient) and (2) the dissipated kinetic energy is efficiently converted to radiation, the decay speed would never exceed $L \propto t^{-1}$. 

\section{SN Ejecta - CSM Interaction Model For SNe Ibn}\label{sec:model}

Hydrodynamic evolution of the SN-CSM interaction, characterized by the forward shock (FS), contact discontinuity (CD), and the reverse shock (RS), is a classical problem with a well-developed solution. The evolution can be described by a self-similar solution \citep{chevalier1982} as long as each of the ejecta and CSM structures is described by a single power-low as a function of radius (which is the case when the reverse shock has not yet been reached to the inner, flat part of the ejecta). Based on the hydrodynamic evolution, the models for thermal X-ray emission from the hot regions behind the FS and RS, as well as the non-thermal emission across wavelengths, have been intensively investigated \citep[e.g.,][for a review]{chevalier2006}.

However, once the problem is on the radiation output in the (thermal) optical emission following the SN-CSM interaction, surprisingly a detailed model is generally lacking. A key issue is how the kinetic energy dissipated at the FS/RS is converted to the thermal radiation energy in the optical wavelengths. A simplified assumption of constant conversion efficiency (which basically assumes full thermalization of the dissipated energy) has been adopted in analytical treatment \citep{chat2012,moriya2013,moriya2014a} which are frequently used to model the LCs of the interaction-powered SNe including SNe Ibn \citep[e.g.,][]{pellegrino2021}. Widely used numerical LC codes also assume full thermalization at the FS and RS \citep{morozova2015,moriya2017,dessart2021}. As we show in the present work, the conversion efficiency should be generally time-dependent \citep[see also][]{chugai2009,chugai2018}, and this effect is important in calculating the LCs for SNe Ibn. 

In the present work, we provide a model which starts with the well-developed formalism describing the FS/RS hydrodynamic evolution and the resulting high-energy radiation. How the high-energy radiation is converted to the (optical) thermal radiation in the interacting system is then considered, which leads to the bolometric (optical) LCs where the conversion efficiency is directly computed within the model framework. Our model is indeed similar to a pioneering work on the SN-CSM interaction LC model formalized by \citet{chugai2001} and then applied to SN Ibn 2006jc \citep{chugai2009} \citep[see also a recent work by][]{tsuna2019}, while there are some difference in detailed treatment. We further aim at investigating the general behavior expected in such a model, and clarifying why and how the model explains the general properties of the SN Ibn LCs. We also investigate how the resulting LC is dependent on the model parameters, which is used to constrain the general properties of SN Ibn ejecta and CSM around them. The model is applied to the LCs of individual SNe Ibn in Section 4. 

\subsection{Models}\label{sec:model-model}

\begin{figure*}[t]
\centering
\includegraphics[width=2.0\columnwidth]{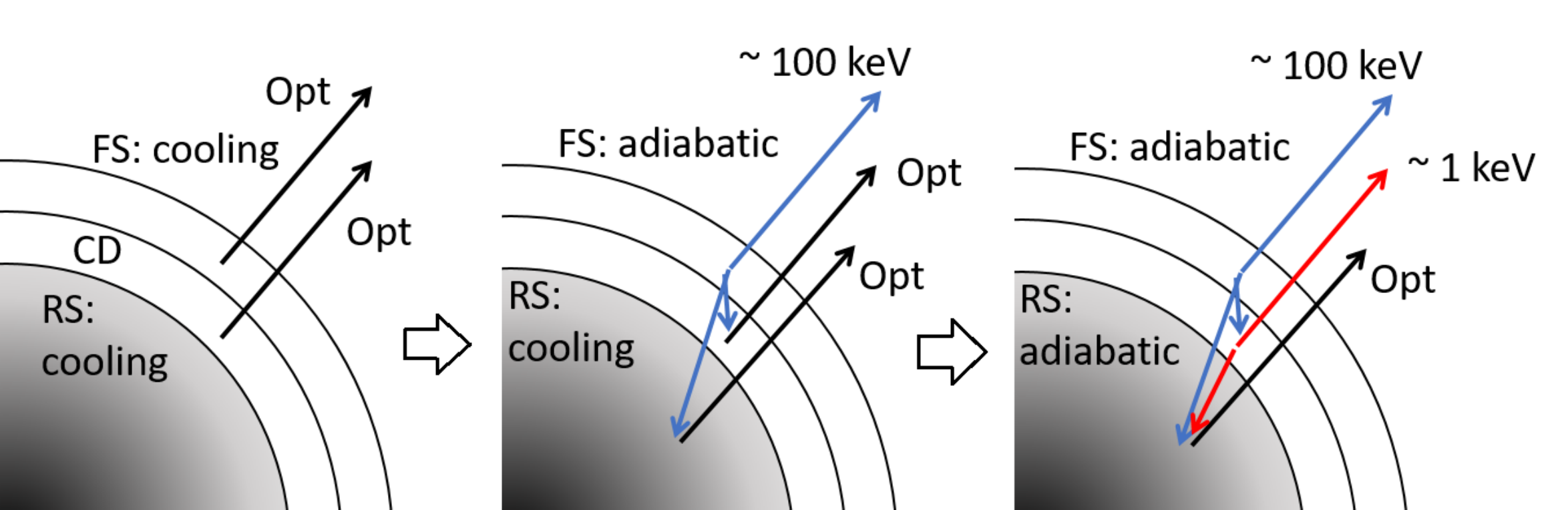}
\caption{A schematic picture of the SN ejecta-CSM interaction. Initially both FS and RS are in the optically-thick cooling regime and the dissipated kinetic energy is immediately converted to the optical photons. Once the FS becomes either in the adiabatic regime or in the optically thin regime (to the 100 keV photons), the dissipated energy is carried by the high-energy photons at $\sim 100$ keV. The inward-directed photons are then converted to the optical photons either in the RS region or in the ejecta, or escape the system as the high-energy photons. As time goes by the RS becomes adiabatic and/or optically thin (to the soft X-ray photons), and the emitted photons at $\sim 1$ keV either escape the system or are converted to the optical photons within the ejecta. 
}
\label{fig:schematic}
\end{figure*}

The configuration considered in the present work is schematically described in Fig. \ref{fig:schematic}. We assume that the SN ejecta are described by a broken power-law in the density structure, where the inner density is constant and the outer part is given as $\rho_{\rm SN} \propto v^{-n}$ \citep[e.g.,][]{chevalier1982,moriya2013}. We fix $n=7$ in the present study, which is within the range expected for an explosion of a compact He star progenitor \citep{chevalier2006,matzner1999}\footnote{We have also investigated the models with $n=10$, and confirmed that our main conclusions would not change.}. With this ejecta structure, the ejecta properties are specified by the ejecta mass ($M_{\rm ej}$) and the kinetic energy ($E_{\rm K}$), which we vary as the input parameters. For the CSM structure, we assume a power-law density distribution;
\begin{equation}
\rho_{\rm CSM} = D r^{-s} = 10^{-14} D' \left(\frac{r}{5 \times 10^{14} \ {\rm cm}}\right)^{-s} \ {\rm g} \  {\rm cm}^{-3} \ .  
\end{equation}
Namely, the normalization $D'$ is taken to be unity for $10^{-14}$ g cm$^{-3}$ at $5 \times 10^{14}$ cm. The corresponding mass-loss rate at $5 \times 10^{14}$ cm (0.03 yr before the SN if the mass-loss wind velocity, $v_{\rm w}$, is $1,000$ km s$^{-1}$) is the following; 
\begin{equation}
\dot M \sim 0.05 D' \left(\frac{v_{\rm w}}{1,000 \ {\rm km} \ {\rm s}}\right) \ M_\odot \ {\rm yr}^{-1} \ . 
\end{equation}
The mass-loss rate here ($D'=1$) corresponds to $A_{*} \sim 5,000$ in another frequently-adopted density scale, where $A_{*} = 1$ is for the combination of $\dot M = 10^{-5} M_\odot$ yr$^{-1}$ and $v_{\rm w} = 1,000$ km s$^{-1}$, i.e., for a typical WR star. For both SN ejecta and the CSM, we assume the He-rich composition. 

The SN-CSM interaction creates the forward shock (FS) and reverse shock (RS), separated by the contact discontinuity (CD). The hydrodynamic evolution of the system can be solved analytically, e.g., by the self-similar solution given by \citet{chevalier1982}. We note that the self-similar solution assumes that the two shocks are in the adiabatic regime with a negligible radiation loss, which is not necessarily the case in the high-density situation considered in the present work, especially for the RS. However, in the cooling regime the detailed structure of the shocked regions is not important for the purpose of this work, since only the rate of the dissipation of the kinetic energy matters in such a regime\footnote{Indeed, most of the analytical works of SN IIn LCs adopt the self-similar solution in the adiabatic regime and compute the radiation output assuming it is in the cooling regime \citep[e.g.,][]{chat2012,moriya2013}. \citet{moriya2013} showed that the resulting LC is consistent with the result of numerical (LTE) radiation-hydrodynamic simulations.}. Further, a key regime we will see in the present work is controlled by the FS in the adiabatic regime (e.g., Section 3.2). 

%The post-shock ion temperatures are given as follows; 
%\begin{eqnarray}
%T_{\rm FS} & = & 1.36 \times 10^9 \left(\frac{n-3}{n-s}\right)^2 %\left(\frac{V}{10^9 \ {\rm cm} \ {\rm s}^{-1}}\right)^2 \ {\rm K}, \\
%T_{\rm RS} & = & \left(\frac{3-s}{n-3}\right)^2 T_{\rm FS} \ , 
%\end{eqnarray}
The post-shock ion temperatures in the FS and RS regions are computed using the expansion velocity of the contact discontinuity through the requirement that the ion bulk kinetic energy is converted to the internal energy at the shock, and they are connected by $T_{\rm RS} =  ((3-s)/(n-3))^2 T_{\rm FS} \propto V^2$, where $V$ is the expansion velocity of the contact discontinuity. With $V \sim 10^9$ cm s$^{-1}$ (which is to be obtained by the SN-CSM ejecta dynamics), the typical post-shock (ion) temperature is $\sim 10^9 K$ for the FS and $\sim 10^7 K$ for the RS. In the present work, we assume that the electrons quickly reach to the equipartition with ions, and thus the (initial) post-shock electron temperatures are given by the above estimate. The assumption of the electron-ion equipartition will also be checked with the Coulomb collision time scale \citep[e.g., ][]{chevalier2006}. 
%\begin{equation}
%t_{\rm e} \sim 2.5 \times 10^7 \left(\frac{T_{\rm e}}{10^9 K}\right)^{1.5} %\left(\frac{n_{\rm e}}{10^7 \ {\rm cm}^{-3}}\right)^{-1} \ s \ , 
%\end{equation}
%where subscript $e$ refers electrons. 
As we will see below, the equipartion is generally realized in the situation studied in the present work. 

The photons that are initially produced in each region have the characteristic energy of $\sim 100$ keV (FS; hereafter `FS photons') and $\sim 1$ keV (RS; `RS photons'). The (absorption) opacity ($\kappa$) in each region for these photons is thus dominated by the Compton scattering and photoelectric absorption, respectively. Assuming $\kappa \sim 0.1$ cm$^{2}$ g$^{-1}$ for the former (i.e., in the He-rich composition) and $\sim 60$ cm$^{2}$ g$^{-1}$ for the letter (i.e., the inner-shell electrons of metals, assuming that the metals are not fully ionized in the RS region), we compute the optical depth and diffusion time within each region for these high energy photons. If the diffusion time exceeds the dynamical time scale, we assume that the photons are trapped within each region and would not make any contribution to the optical output. The emissivity for these high energy photons is computed with the cooling function presented by \citet{chevalier2006}, with the upper limit set by the total kinetic power dissipated at the corresponding shock. 

For the FS photons, we assume that a half of the luminosity always escapes from the system as the high-energy photons (as long as the corresponding diffusion time is shorter than the dynamical time scale). The remaining half is assumed to be directed inward. We judge the characteristic spectral energy of the photons escaping from the FS region to the RS region by the following criteria; (1) the cooling time scale is shorter than the dynamical time scale (i.e., in the cooling regime), and (2) the optical depth to the Compton scattering for the high-energy photons exceeds unity. If these two criteria are satisfied, we assume that the FS region is cooled down and (half of) the energy is carried away from the system as optical photons (the situation we call the optically-thick cooling regime, noting that it is optically thick for the high-energy photons, not for the optical photons). 

In case the inward radiation from the FS is of high energy (middle panel of Fig. \ref{fig:schematic}), we further compute the Compton optical depths for these photons within the RS and the freely-expanding unshocked ejecta, and convert these to the fractions of the FS high-energy photons absorbed either in the RS region or in the ejecta. These radiation inputs are then added to the $\sim 1$ keV photons originated in the RS region and the optical photons in (the outer layer of) the ejecta. respectively. 

The treatment of the X-ray photons originated in the RS region follows a similar manner. The total energy input from the RS region is a sum of the one computed by the cooling function for the RS region and the one deposited by the FS photons (see above). If the RS region is judged to be in the optically-thick cooling regime, it is assume that the RS region is quickly cooled down and all the energy input to the RS region is converted to optical photons. Otherwise, the radiation from this region escapes as X-ray photons ($\sim 1$ keV) from the RS region, as long as the diffusion time scale is shorter than the dynamical time scale (right panel of Fig. \ref{fig:schematic}). A half of the X-ray energy is assumed to be directed inward; the fraction of the energy deposited to the unshocked ejecta (thus the contribution to the optical luminosity) is computed in the same manner as the FS photons (but with the opacity at $\sim 1$ keV). 

The above procedures yield the optical luminosity as a sum of the direct contributions from the FS and RS regions (if these regions are in the optically-thick cooling regime) and the energy input to the unshocked ejecta by the high-energy photons originally emitted at the FS and RS regions. Finally, we compute the diffusion time scale for the optical photons within the FS/RS regions and the unshocked CSM, and this optical luminosity is set as the observed optical (bolometric) luminosity if the diffusion time scale is shorter than the dynamical time scale. While it is not relevant for the purpose of this work, we also compute the high-energy radiation luminosity at $\sim 1$ keV (RS) and $\sim 100$ keV (FS), taking into account the optical depths of the unshocked CSM for these high-energy photons.

\subsection{General Properties}\label{sec:model-general}

\begin{figure*}[t]
\centering
\includegraphics[width=\columnwidth]{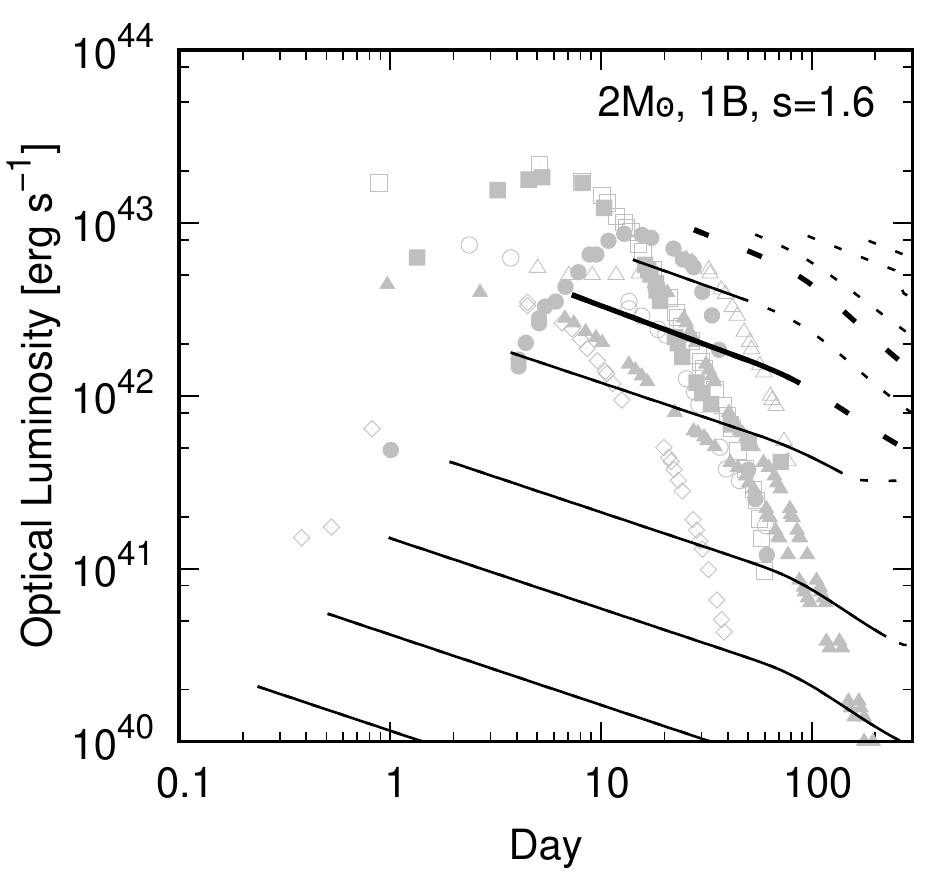}
\includegraphics[width=\columnwidth]{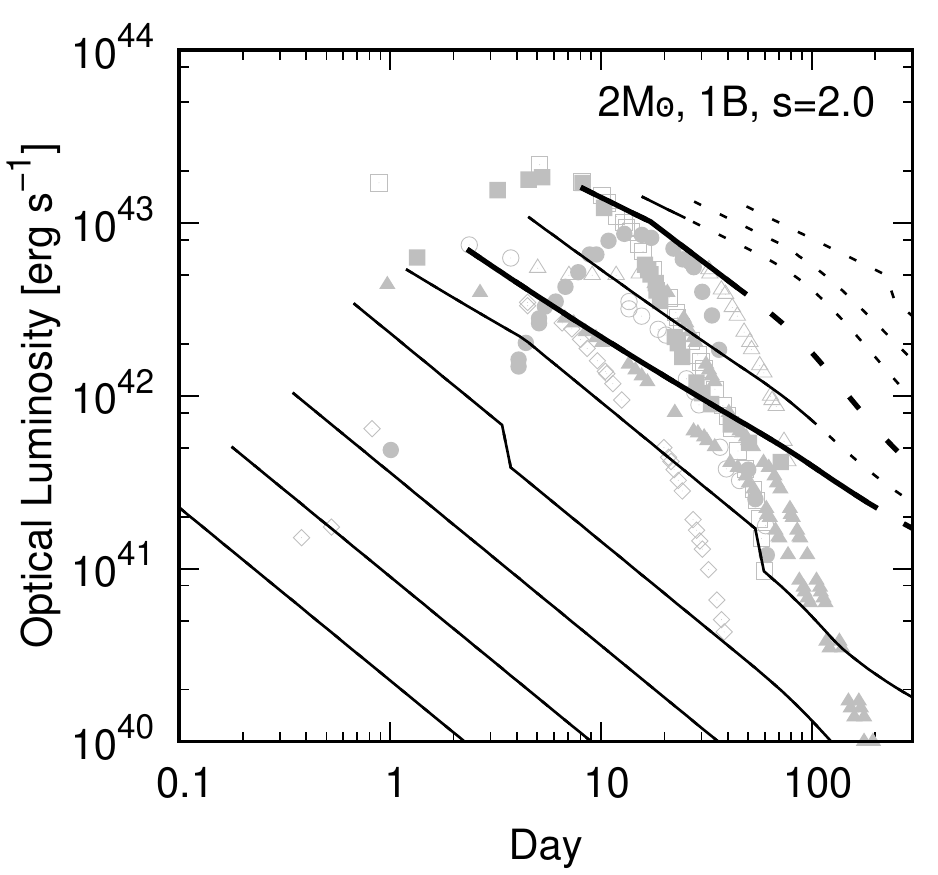}\\
\includegraphics[width=\columnwidth]{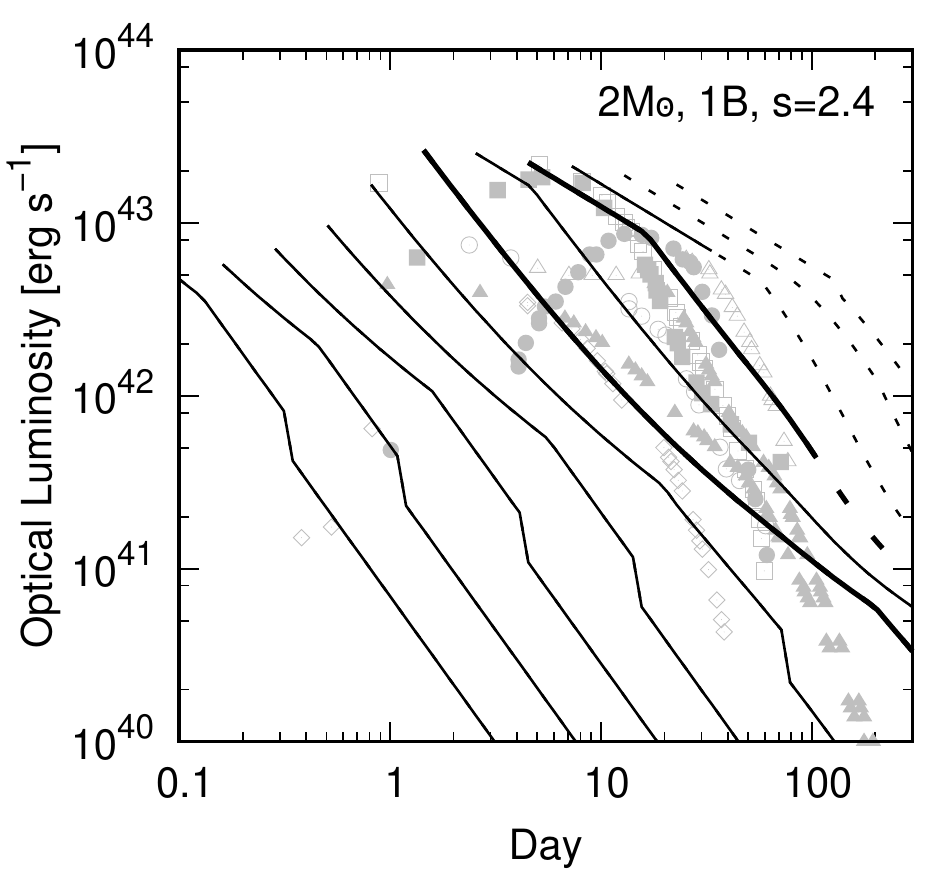}
\includegraphics[width=\columnwidth]{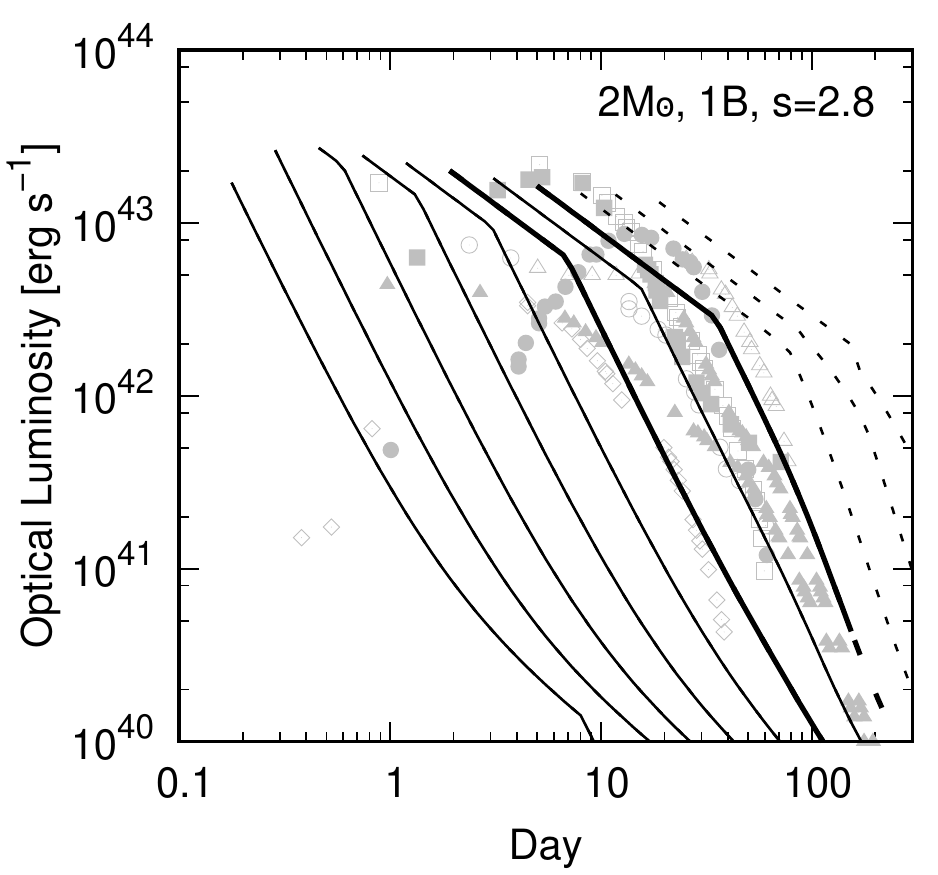}
\caption{The light curves for the SN-CSM interaction model for SNe Ibn. The ejecta mass and the kinetic energy are set to be $2 M_\odot$ and $10^{51}$ erg, which are typical for SNe Ibc. The four cases are shown changing the slope of the CSM density distribution ($\rho \propto r^{-s}$): $s = 1.6$, $2.0$, $2.4$ and $2.8$. The CSM density scale, $D'$, is varied from $0.03$ to $32$ with a step of a factor of two each. The models with $D' = 1$ and $4$ are shown by the thick lines. After the reverse shock is estimated to reach to the inner (flat) ejecta, an assumption in the model is not justified anymore, and the LCs there are shown by dashed lines (for large $D'$). The `starting time' in the model LCs is set by the diffusion time scale for the optical photons; we attribute it to the peak luminosity. It is expected that the SN is in the rising phase before this epoch, but the diffusion process to construct the LC in this rising phase is not solved in the present work. 
}
\label{fig:reference}
\end{figure*}

\begin{figure*}[t]
\centering
\includegraphics[width=0.6\columnwidth]{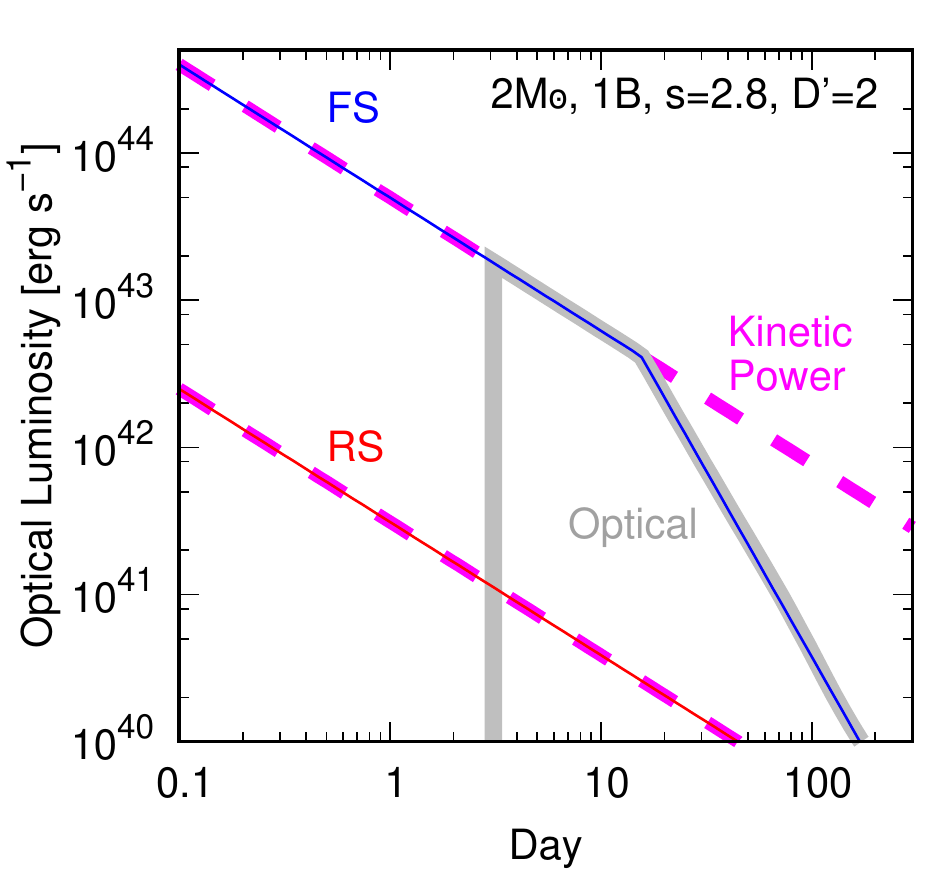}
\includegraphics[width=0.6\columnwidth]{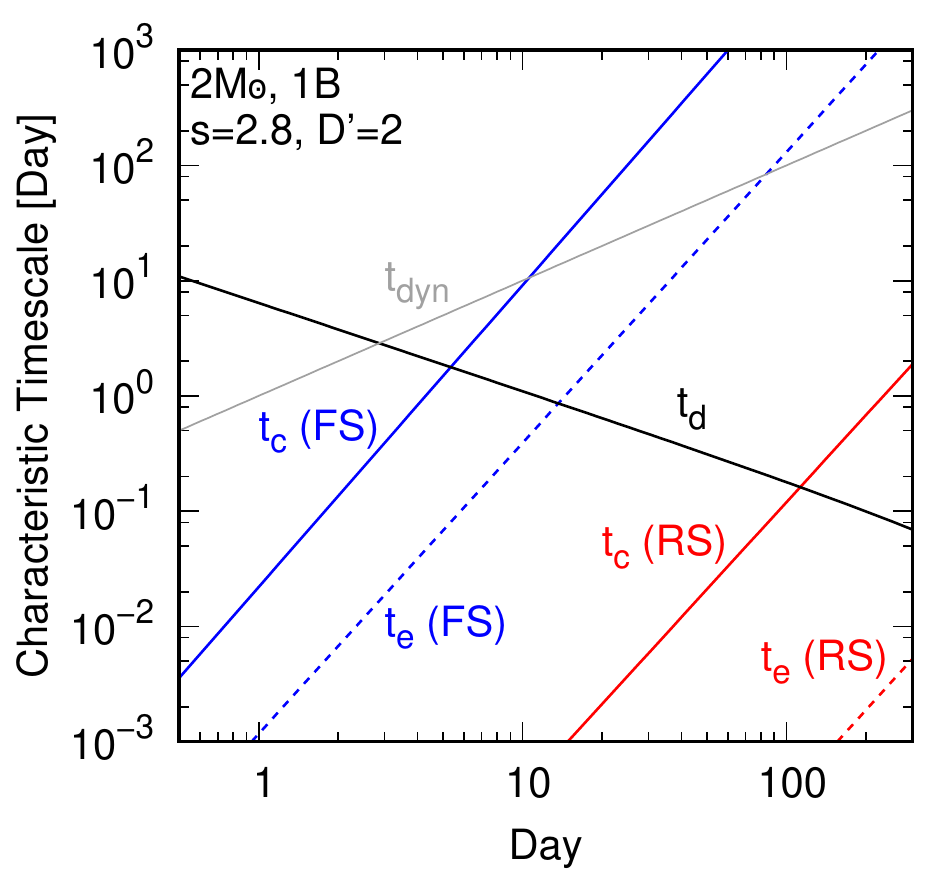}
\includegraphics[width=0.6\columnwidth]{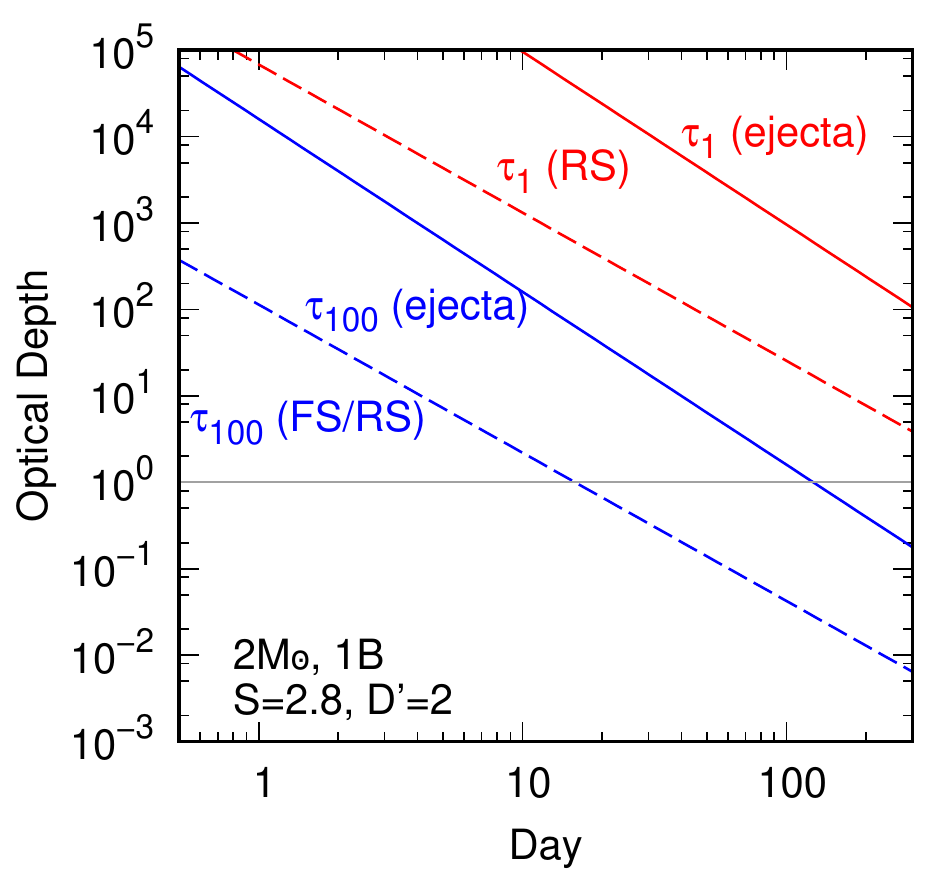}\\
\includegraphics[width=0.6\columnwidth]{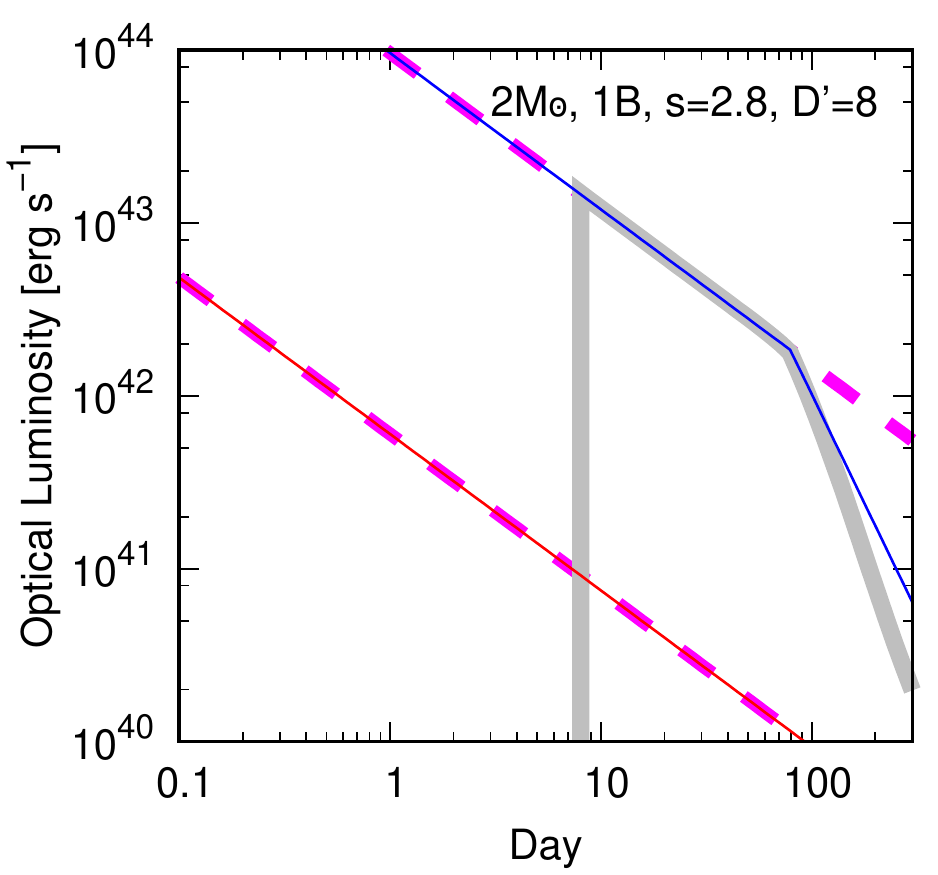}
\includegraphics[width=0.6\columnwidth]{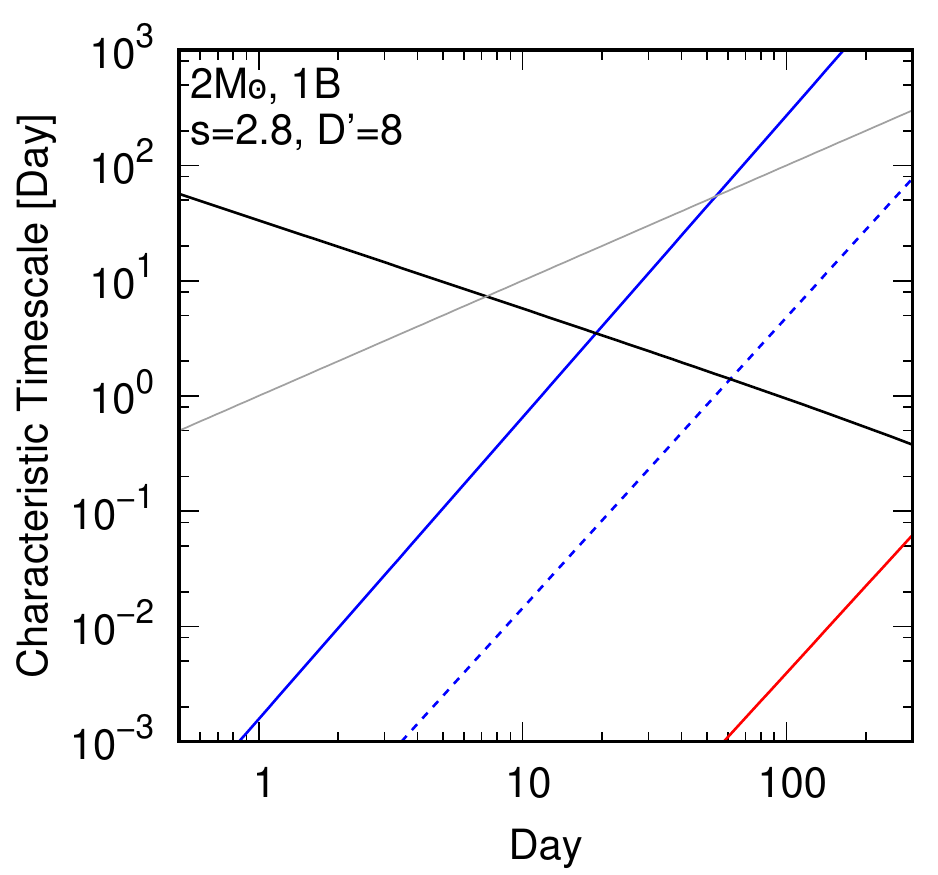}
\includegraphics[width=0.6\columnwidth]{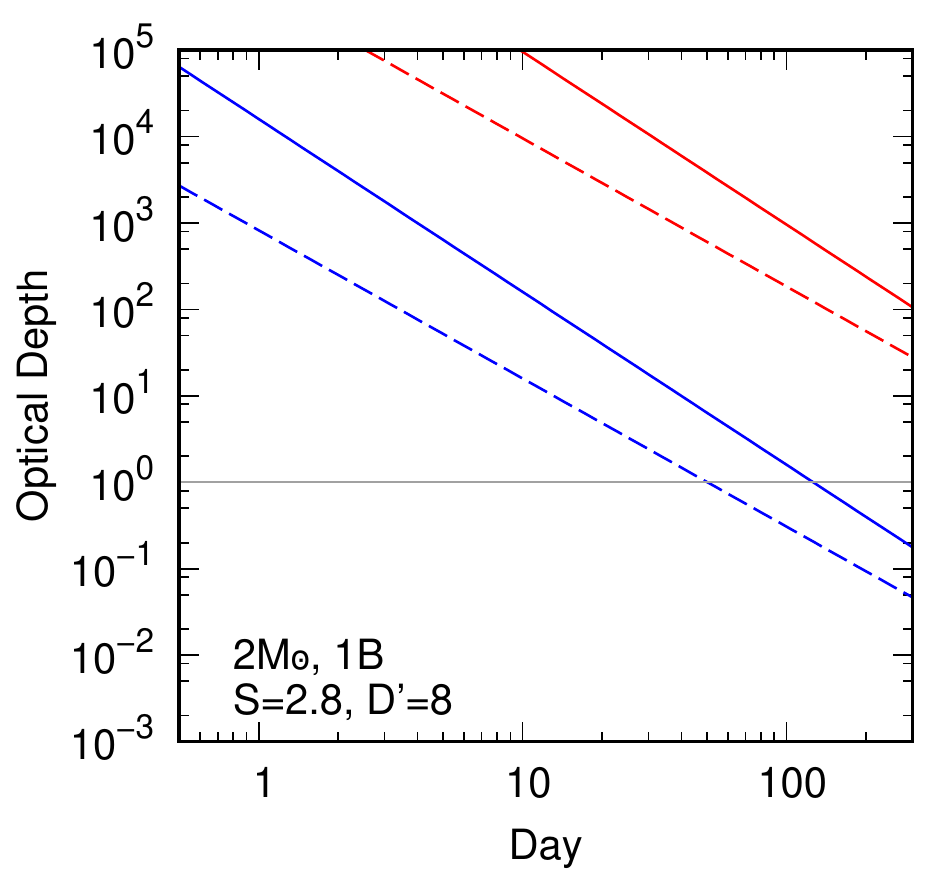}\\
\includegraphics[width=0.6\columnwidth]{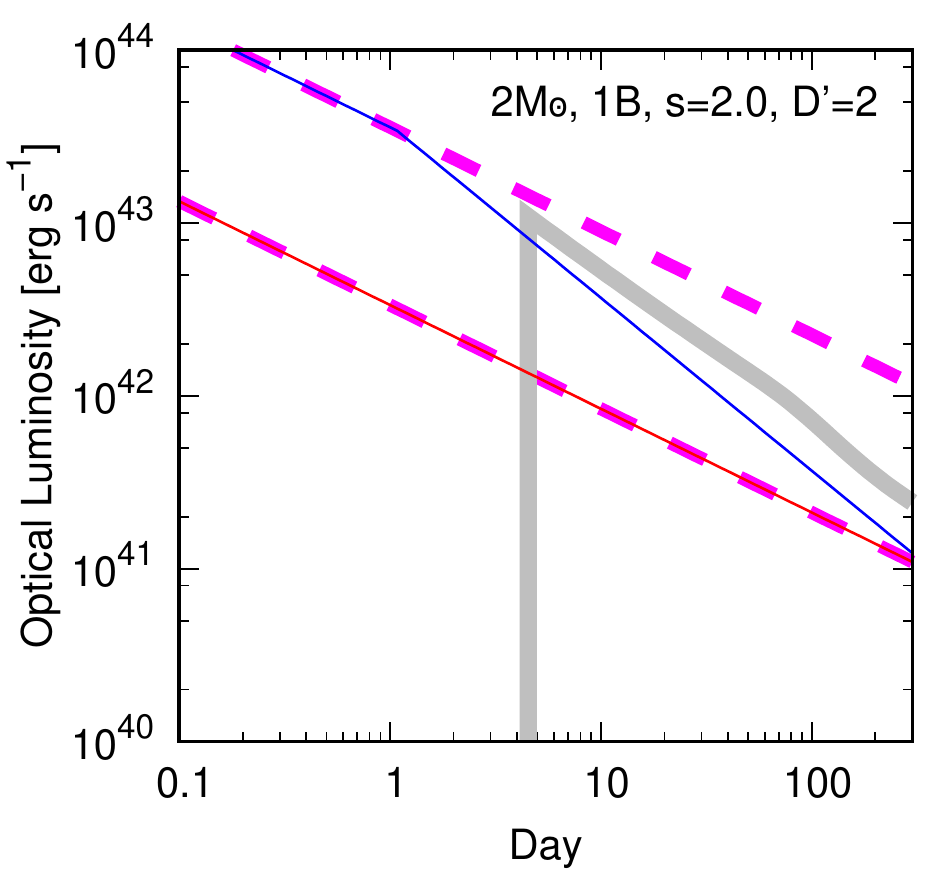}
\includegraphics[width=0.6\columnwidth]{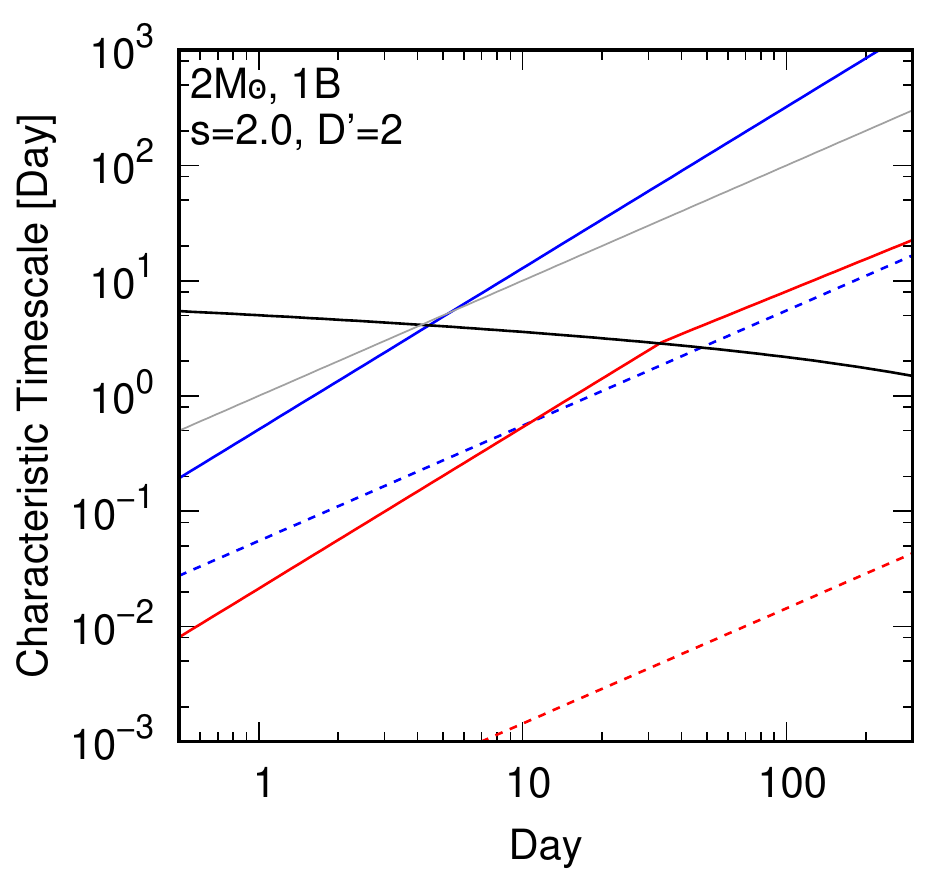}
\includegraphics[width=0.6\columnwidth]{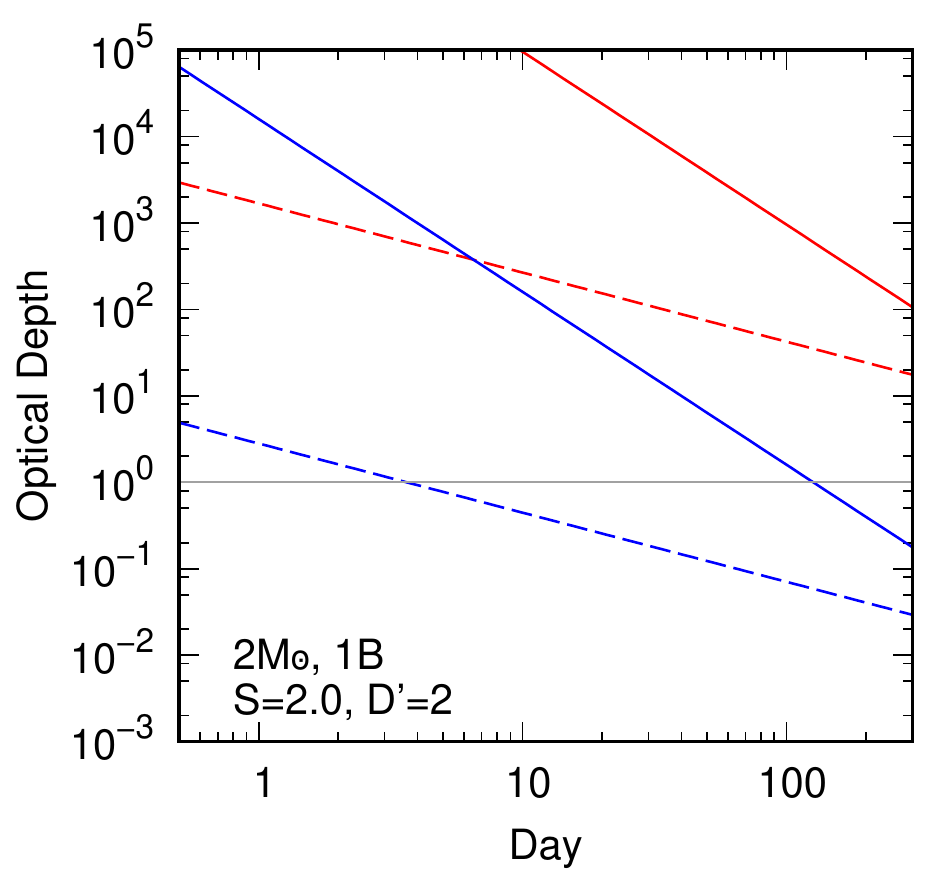}\\
\includegraphics[width=0.6\columnwidth]{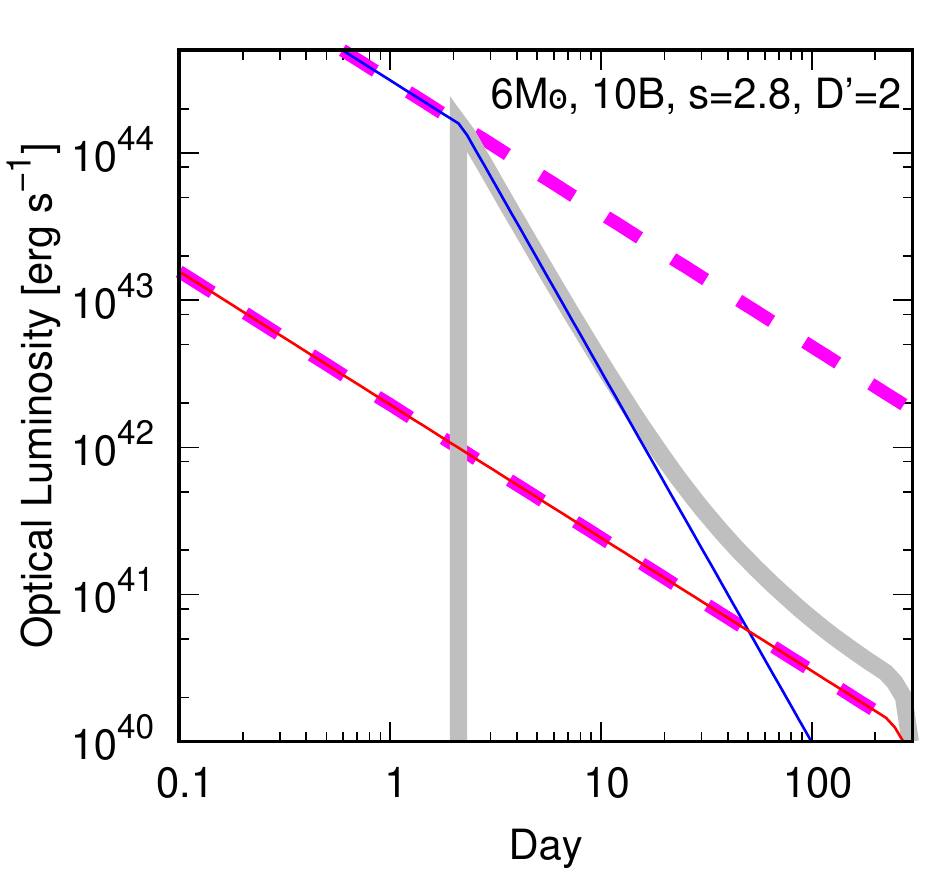}
\includegraphics[width=0.6\columnwidth]{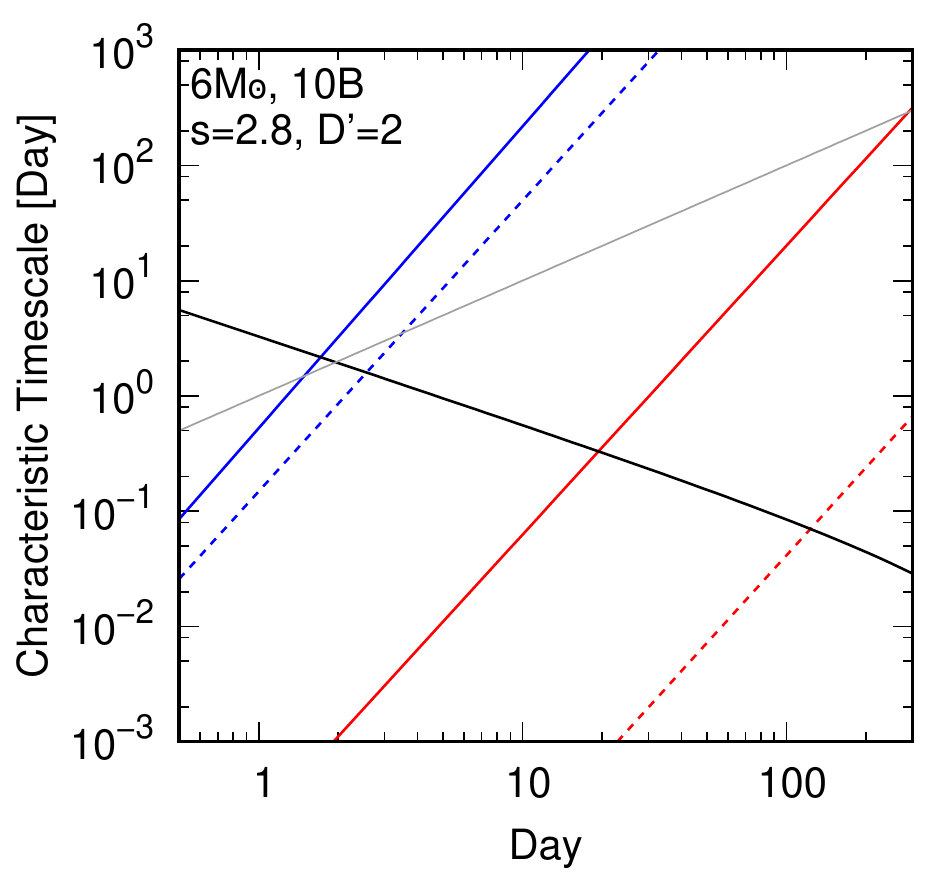}
\includegraphics[width=0.6\columnwidth]{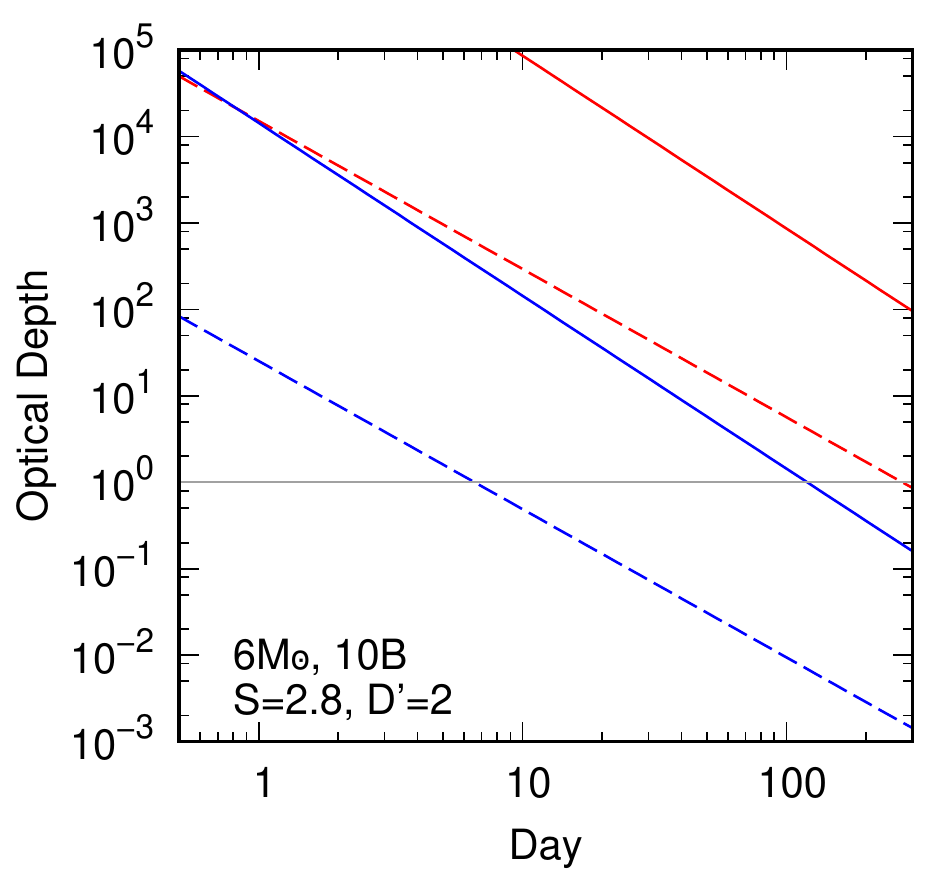}
\caption{The behaviours of the selected four models (from the top to bottom panels). The left panels show the LCs deconvolved into the contributions from the FS (blue-solid) and RS (red-solid) regions. The kinetic energy dissipation rate in each region is shown by the magenta-dashed lines. The resulting optical (bolometric) LC is shown by the gray line. The middle panels show characteristic time scales; dynamical time scale ($t_{\rm dyn}$; gray), diffusion time scale of the optical photons ($t_{\rm d}$; black), cooling time scale ($t_{\rm c}$; solid) and electron-ion euqipartition time scale ($t_{\rm e}$; dotted) in the FS (blue) and RS (red) regions. The right panels are for the characteristic optical depths for the following; the photons originally emitted at the FS ($\sim 100$ keV) traveling either in the FS or RS region ($\tau_{100}$; blue-dashed; note that the optical depths in the FS and RS are similar in the present work), the photons emitted at the RS ($\sim 1$ keV) traveling within the RS ($\tau_{1}$; red, dashed), those emitted either in the FS (blue-solid) or RS (red-solid) traveling into the ejecta. 
}
\label{fig:component}
\end{figure*}

Figure \ref{fig:reference} shows examples of the model LCs, in which the CSM properties ($s$, $D'$) are varied for the fixed ejecta properties ($M_{\rm ej} = 2 M_\odot$ and $E_{\rm K} = 10^{51}$ erg $= 1 $ Bethe). The luminosity at a given epoch is higher for larger $D'$ (i.e., denser CSM) and the evolution is quicker for larger $s$ (i.e., steeper CSM density distribution), both of which are generally expected in the SN-CSM interaction. To account for the peak luminosity of SNe Ibn, $D' \sim 0.5-5$ is required in this scenario. The characteristic slope in the late-time rapid decay phase ($L \propto t^{-3}$; see Section 2) is explained by a steep CSM density gradient ($s \sim 3$). The `starting time' in each model corresponds to the peak luminosity, and it is expected that the LCs are in the rising phase before that epoch (but it is not solved in the present study). The peak date is delayed as $D'$ increases, as expected. We note that while a larger peak luminosity is reached for a model with later peak date for the CSM with relatively flat CSM distribution ($s \sim 2$ or flatter) as seen in many numerical simulations for the SN IIn LCs \citep[e.g.,][]{moriya2013}, it is not necessarily the case for a steep CSM density distribution (e.g., $s \sim 3$) because of a larger energy input at earlier date (with steeply increasing CSM density inward). We also note that the initial, post-peak LC decay rates for the flat CSM model ($s = 1.6$) and the steep CSM model ($s = 2.8$) are not very different, which might naively be against physical intuition; the (immediate) post-peak phase is indeed powered by different mechanisms in these two extreme cases, where it is (already) the adiabatic FS in the former while it is (still) the cooling FS in the latter. This behavior is addressed below in more detail. 

Figure \ref{fig:reference} shows that a typical SN Ibc ejecta model exploding within a high-density CSM with a steep density gradient ($s \sim3 $) explains typical (homogeneous) LC evolution of SNe Ibn. To further clarify the model behaviors, Fig. \ref{fig:component} shows the contributions from the FS and RS regions, characteristic time scales, and optical depths, for a few selected models. 

The model with $M_{\rm ej} = 2 M_\odot$, $E_{\rm K} = 10^{51}$ erg, $s=2.8$ and $D' = 2$ is our reference model for SNe Ibn, which is shown in the top panels of Fig. \ref{fig:component}. In this model, the RS is always (up to $\sim 100$ days) in the optically-thick cooling regime ($t_{\rm c} (RS) < t_{\rm dyn}$ for the cooling and dynamical timescales, and $\tau_{1} (RS) > 1$ for the optical depth for $\sim 1$ keV photons within the RS region). Therefore, the X-ray photons originally emitted within the RS region will be quickly converted to optical photons. Despite the high efficiency of converting the dissipated kinetic energy to the optical photons at the RS region, the intrinsic power is generally negligible as compared to that in the FS region (see Section 4.1). The contribution from the RS to the optical radiation becomes important only after $\sim 100$ days. The FS is in the cooling regime in the first $\sim 10$ days, and in the adiabatic regime afterword. In the early cooling phase, the optical depth for the high-energy photons (at $\sim 100$ keV) is above the unity within the FS, and thus the optical luminosity traces the rate of the kinetic energy dissipation. After $\sim 10$ days when the FS becomes adiabatic, the optical depth for the FS photons is below unity; a fraction of the FS photons is then absorbed in the RS. The remaining high-energy photons is all absorbed in the ejecta up to $\sim 100$ days. Therefore, ultimately all the energy input at the FS (subtracting another half of the power which is assumed to be directed outward and escape the system) is converted to the optical luminosity in the time window of interest in the present work. We note that the assumption of the electron-ion equipartition in the FS region is not justified after $\sim 100$ days, but this is basically beyond the phase of the observations for which the data have been taken for the sample of SNe Ibn studied here; further, in such a late phase, the contribution from the RS starts dominating the optical output, for which the equipartition is still realized. 

The denser CSM ($D' = 8$, shown in the second panels from the top in Fig. \ref{fig:component}) leads to the properties similar to the reference model. The difference from the reference model ($D' = 2$) can be understood in a straightforward manner; it is bright thanks to the high density, and the transition from the post-peak slow (cooling) phase to the rapid decay (adiabatic) phase is delayed for the same reason. The longer diffusion time scale for the optical photons leads to the peak date reached later than in the reference model. 

The effect of the CSM density slope is highlighted in the third panels from the top in Fig. \ref{fig:component}. Here, $s=2$ is adopted (i.e., the steady-state mass loss). Due to the flatter CSM density distribution, the LC evolution is generally slower than in the reference model. The peak luminosity is comparable to the reference model; $D'$ is normalized at $5 \times 10^{14}$ cm, which is the physical scale of the shock wave expansion around the peak date (and $D'$ is taken to be identical between this model and the reference model). The behavior of the RS contribution is essentially the same with the reference model; it is always in the optically-thick cooling regime. The relative contribution of the RS to the FS is larger in this model, which is generally expected for the flatter CSM distribution. 

One important, qualitative difference is in the power input by the FS. The FS is already in the adiabatic phase at the peak date, and thus the entire LC evolution is basically controlled by the FS energy input in the adiabatic phase. In this phase, the decline in the LC is slower for the flatter CSM distribution. As a combination of these two effects, the post-peak decline rate is indeed similar to that in the reference model, despite very different regimes for the post-peak emission. 

Finally, how the SN ejecta properties affect the LC is (partly) addressed in the bottom panels. The model here assumes the SN properties corresponding to the so-called broad-lined SNe Ic (SNe Ic-BL) or hypernovae (HNe), sometimes associated with a Gamma-Ray Burst (GRB) \citep{galama1998,iwamoto1998,woosley2006}. The model here represents an energetic SESN explosion with $M_{\rm ej} = 6 M_\odot$ and $E_{\rm K} = 10^{52}$ erg. Due to the high shock velocity, the model results in a large kinetic energy dissipation and initially a bright emission.The high velocity also leads to a large shock radius and thus a low CSM density; it thus results in a quick transition from the cooling to adiabatic regime for the FS, and it is already adiabatic at the peak date. It thus lacks the post-peak slow decay phase and shows a quick decay already at the beginning. The RS is still in the cooling regime even for this model, and thus the main power source changes from the FS to the RS at $\sim 50$ days, showing the flattening in the LC afterword. It thus results in a qualitatively different LC shape from the reference model. 

\begin{figure}[t]
\centering
\includegraphics[width=\columnwidth]{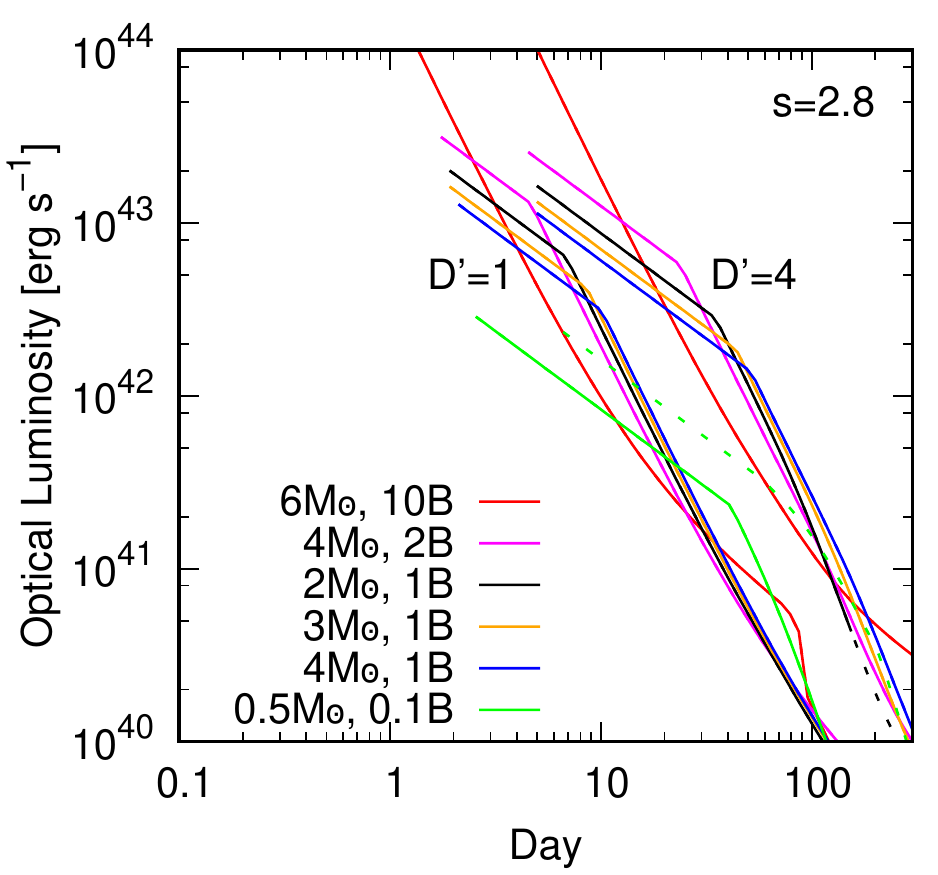}
\caption{Dependence of the interaction-powered SN Ibn LCs on the ejecta properties. The CSM density distribution is fixed to be $s=2.8$, and the CSM density parameter, $D'$, is shown for two cases ($D' = 1$ and $4$). }
\label{fig:ejecta}
\end{figure}

The dependence on the ejecta properties is further shown in Fig. \ref{fig:ejecta}. As long as $M_{\rm ej}$ is a few $\times M_\odot$ and the kinetic energy is $\sim 10^{51}$ erg, the general behavior of the LC is not sensitively dependent on the ejecta properties; they all show the characteristic power-law LC with a break marking a transition from the slow to fast decay phase. This case covers models with relatively massive ejecta, e.g., $\sim 4 M_{\odot}$ (or even larger, while it is not shown here; see Section 5.2). As a general behavior, the luminosity at the slow decay phase tends to be higher for models with larger $E_{\rm K} / M_{\rm ej}$ (i.e., higher velocity) followed by the earlier transition to the fast decay phase.  Interestingly, they marge in the fast decay phase; higher velocity leads to a larger shock radius and thus lower CSM density, and these two effects nearly cancel to result in the similar luminosity.  

This homogeneous nature in the LCs is no more realized if the ejecta properties are substantially outside the range expected for canonical SESNe. Fig. \ref{fig:ejecta} shows two extreme cases; very energetic (and massive) ejecta corresponding to SNe Ic-BL (or GRB-SNe) \citep[see][and references therein]{woosley2006}, and weak SNe with low energy and small ejecta mass corresponding to the so-called ultra-stripped envelope SNe (USSNe), electron-capture SNe, or their variants, for which the observed candidates have been reported recently \citep[e.g.,][and references therein]{de2018,de2021,hiramatsu2021,nakaoka2021}. The former (energetic model) has a very rapid decay from the beginning, a reason of which has been already addressed above. The latter (the low-energy and low-mass SNe) indeed shares the general LC behavior, but the luminosity is lower than the canonical cases in the post-peak slow decay phase, due to the lower kinetic energy content. This model reaches only $\sim 10^{42}$ erg s$^{-1}$, and thus it could explain at most only the most faint SN Ibn population. 

Robustness of constraining the ejecta properties requires further discussion. This is a general problem for SNe powered by the SN-CSM interaction; for example, the problem of uniqueness of ejecta parameters for SNe IIn has been discussed by \citet{chandra2015} and \citet{chugai2021}. The three physical parameters ($M_{\rm ej}$, $E_{\rm K}$, and $D'$) are degenerate to produce the same LC for fixed ejecta and CSM slopes ($n$ and $s$), in each limit of the cooling and adiabatic FS regimes \citep[see, e.g.,][for the cooling regime applied to SNe IIn]{moriya2014a}. It is possible to introduce additional constraint to solve a part of the degeneracy and place separate constraints on the ejecta properties ($M_{\rm ej}$ and $E_{\rm K}$) and the CSM property ($D'$). This can be done by adding information on the velocity from spectra \citep[e.g.,][]{chugai2009}. In the present model, this separation between the ejecta and CSM properties is possible by using the LC evolution covering the cooling and adiabatic FS regimes (e.g., the transition date introduces an additional constraint to the problem). Further solving the degeneracy to derive $M_{\rm ej}$ and $E_{\rm K}$ separately is problematic. Indeed, in the idealized situation of applying the self-similar solution for two power-law segments ($n$ and $s$), it is in principle impossible; the same power-law density distribution of the outer ejecta can be realised by different combinations of ($M_{\rm ej}$, $E_{\rm K}$). Namely, for given $M_{\rm ej}$ one can find $E_{\rm K}$ that can reproduce the identical LC, and thus one can only define a series of ($M_{\rm ej}$, $E_{\rm K}$) from the LC analysis. In the discussion above on the allowed range of the ejecta mass, we take the opposite approach; we first adopt a relation between $M_{\rm ej}$ and $E_{\rm K}$ motivated either theoretically or observationally, and then investigate whether each of the adopted combinations can explain the SN Ibn LCs. We believe that the rough constraint derived by this approach is sufficient for the purpose of the present work; we plan to investigate a possibility of using the velocity evolution and spectral appearance to further constrain the ejecta properties in the future.

\section{Application to a Sample of SNe Ibn}\label{sec:appliaction}

\subsection{Key Behaviours}\label{sec:application-key}

The high optical luminosity of SNe Ibn, which is probably powered by the SN-CSM interaction, suggests that the CSM is very dense. In such a system, at least initially the energy dissipated at the FS and RS should be (nearly) fully converted to the optical radiation (i.e., both FS and RS are in the optically-thick cooling regime). The luminosity is thus estimated by a simple formula frequently adopted for SNe IIn; 
\begin{equation}
L \propto \rho_{\rm CSM} R^2 V^3 \propto R^{2-s} V^3 \ .
\end{equation}
Note that the luminosity from the FS dominates over that from the RS; the ratio of the dissipated power is $L_{\rm FS} / L_{\rm RS} \propto \rho_{\rm FS} / \rho_{\rm RS} (V_{\rm FS}/V_{\rm RS})^3 \propto (4-s)(n-3)^2/(n-4)/(3-s)^2$, and the ratio is $\sim 4$ for $s=0$, $\sim 10$ for $s=2$, or even exceeds $\sim 100$ for $s > 2.7$. 

For the CSM density decreasing outward, first the FS becomes adiabatic due to the lower density than for the RS. The energy input from the FS still dominates for the reason mentioned above. The RS and the unshocked ejecta are still optically thick to the FS photons, and most of the (inward) radiation emitted originally as high-energy photons at the FS will be converted to optical photons. In this situation, the luminosity is estimated as follows; 
\begin{equation}
L \propto T_{\rm FS}^{0.5} M_{\rm FS} \ \rho_{\rm CSM} \propto V R^{3-2s} \ ,
\end{equation}
where $M_{\rm FS}$ is the CSM mass swept by the FS. 

It is then possible to evaluate the power-law behavior of the optical (bolometric) LC in the two phases ($L \propto t^{\beta}$), by inserting the (self-similar) dependence of the evolution of the shock radius on $n$ (ejecta density slope) and $s$ (CSM density slope). For $n=7$ and $s=2$, $R \propto t^{(n-3)/(n-s)} = t^{0.8}$ and thus $V \propto t^{-0.2}$ \citep{chevalier1982}. This results in $\beta \sim -0.6$ in the early (cooling) phase and $\beta \sim -1$ in the late (adiabatic) phase, where the latter dose not depends on $n$. This estimate demonstrates that the observed characteristic behavior, $\beta \sim -1$ in the post peak phase followed by $\beta \sim -3$ in the steep decay phase, is not consistent with the standard CSM distribution ($\rho_{\rm CSM} \propto r^{-2}$) created by the steady-state mass loss, irrespective of the emission processes (either the radiative/cooling or adiabatic). 

A straightforward solution is obtained if $s \sim 3$. In this case, the shock expands almost at a constant velocity, and thus $\beta \sim -1$ in the early phase and $\beta \sim -3$ in the late phase. Given the rapid decrease in the CSM density outward, the transition from the cooling regime to the adiabatic regime is indeed well motivated by the underlying physics, and it does take place in the situation considered in the present work (Section 3). This simple picture explains the general (and homogeneous) LC evolution of SNe Ibn very well. 

As another possibility, one might be tempted to introduce a non-smooth CSM distribution where the inner part is flat and the outer part is steep, under the assumption that the optical luminosity traces the total kinetic power dissipation rate. For the late-phase steep decay, it will require $s \sim 5$. This will then inevitably lead to the adiabatic FS rather than the cooling FS due to the rapidly decreasing density, and the underlying assumptions become self-inconsistent. While we do not reject a possibility of the possible change in the CSM density gradient (see next Section), we conclude that the steep decay phase must be powered by the FS in the adiabatic regime and thus $s \sim 3$ at least in the (relatively) outer CSM density distribution.

\subsection{Individual Objects}\label{sec:individual}

\begin{figure*}[t]
\centering
\includegraphics[width=0.65\columnwidth]{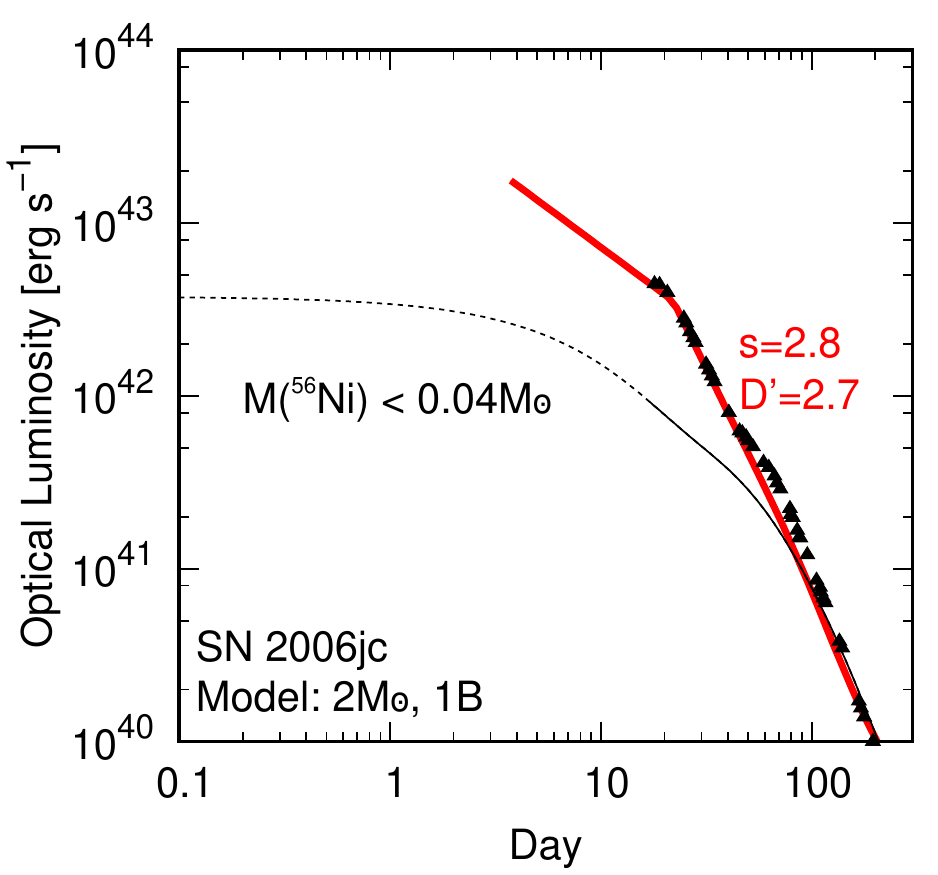}
\includegraphics[width=0.65\columnwidth]{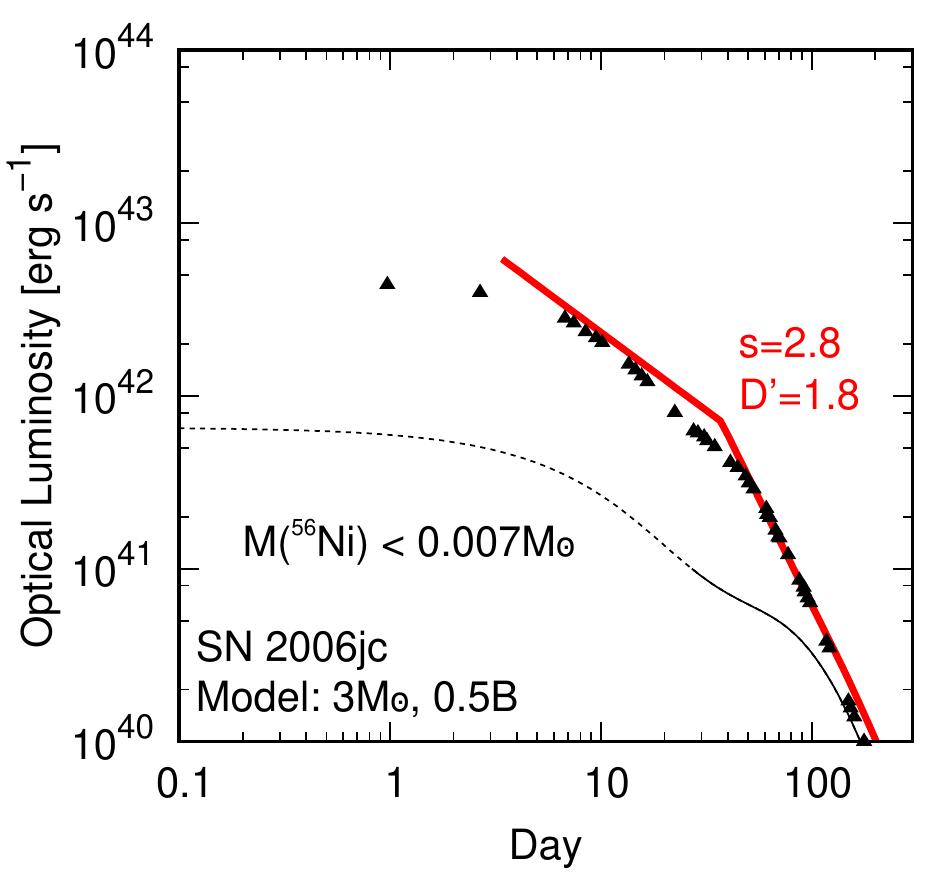}\\
\includegraphics[width=0.65\columnwidth]{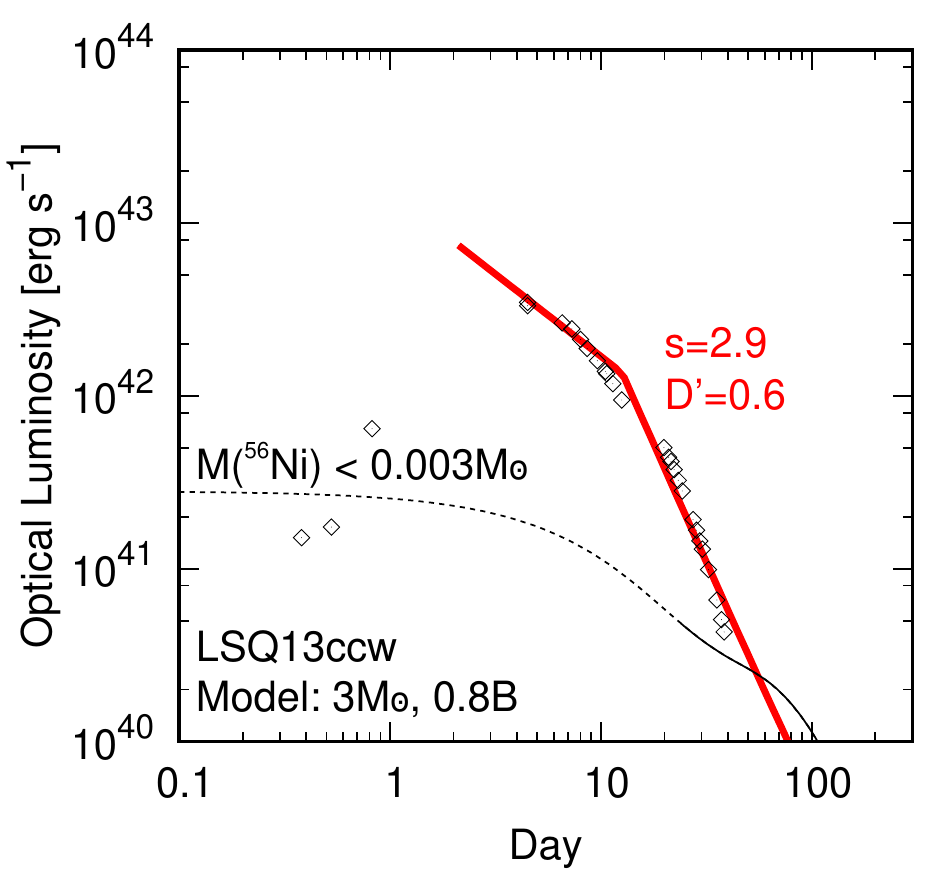}
\includegraphics[width=0.65\columnwidth]{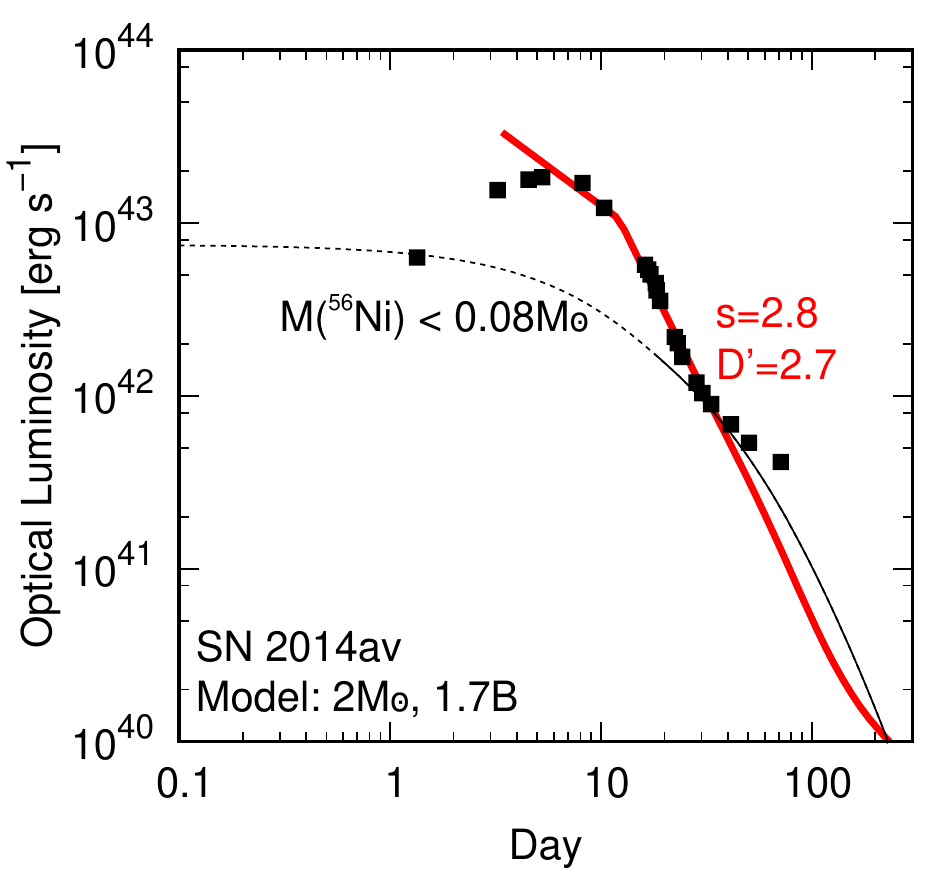}\\
\includegraphics[width=0.65\columnwidth]{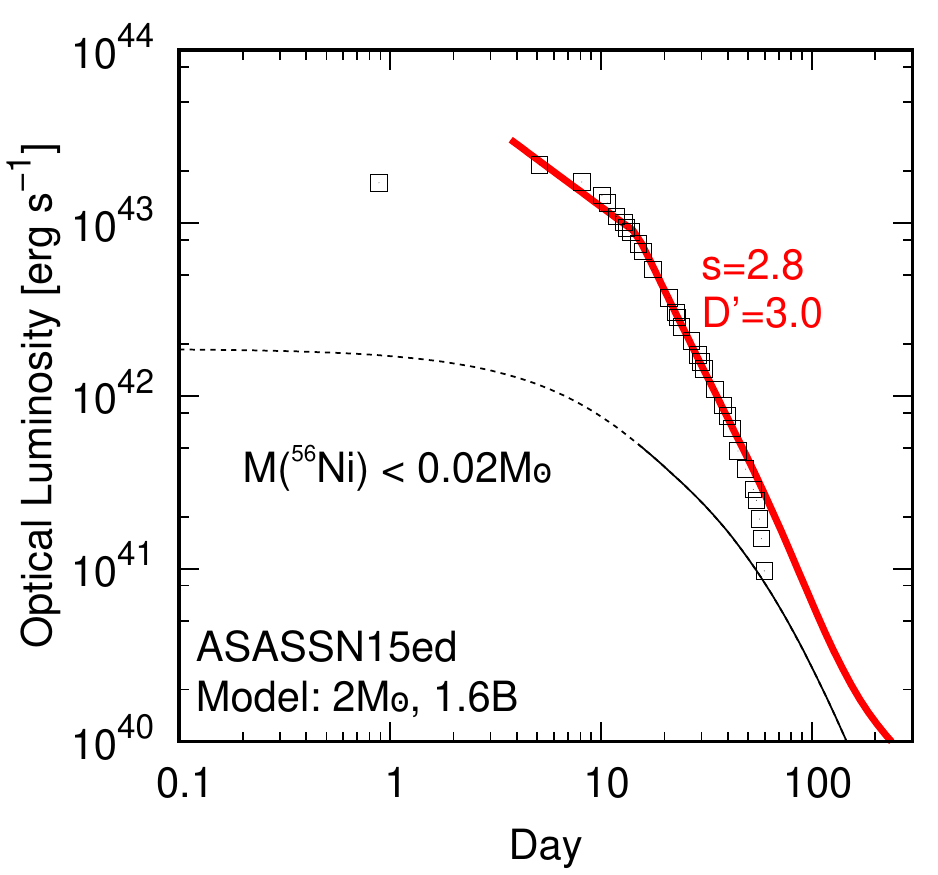}
\includegraphics[width=0.65\columnwidth]{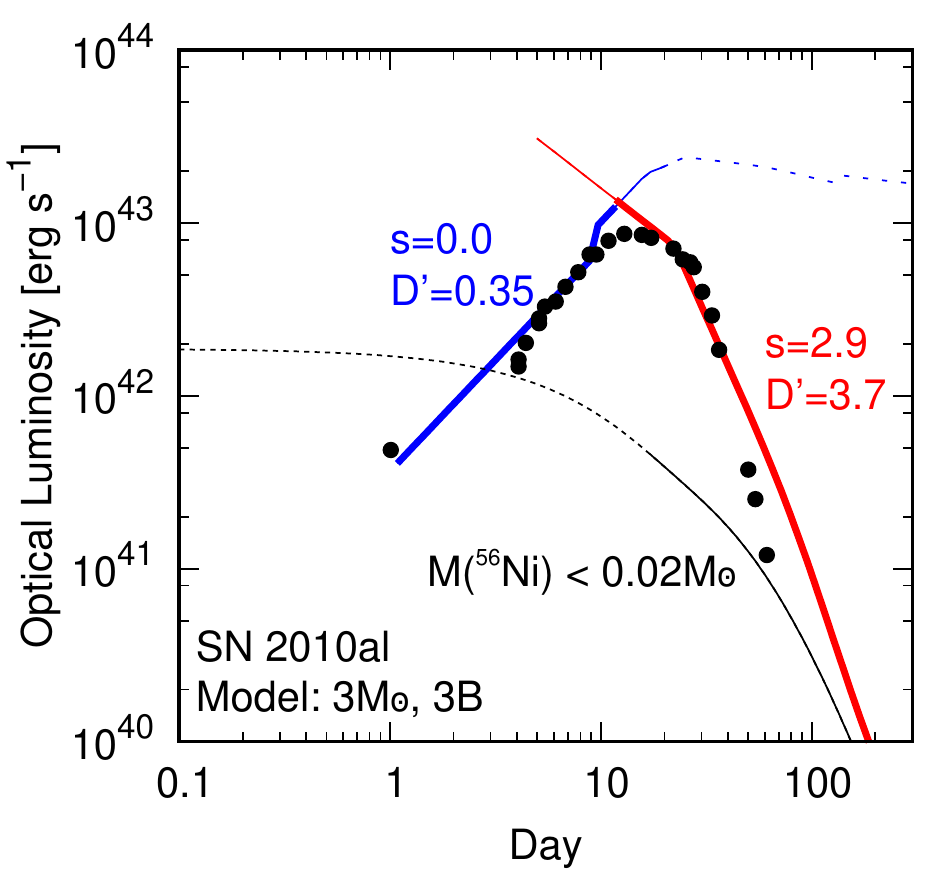}\\
\includegraphics[width=0.65\columnwidth]{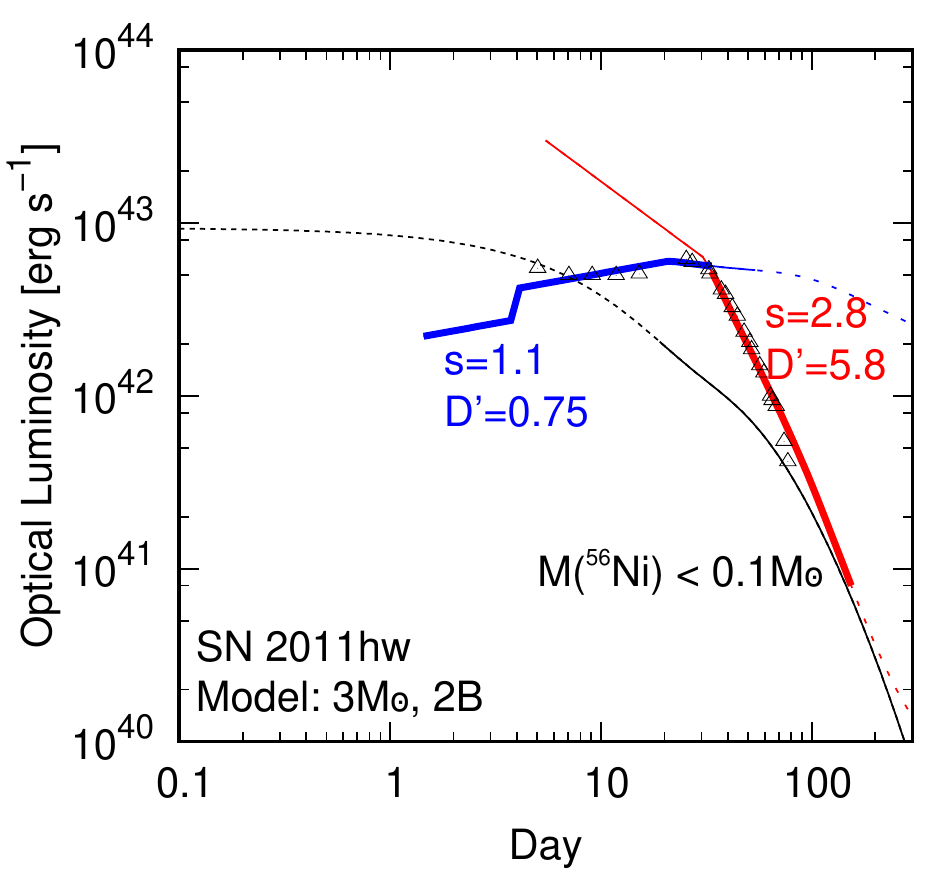}
\includegraphics[width=0.65\columnwidth]{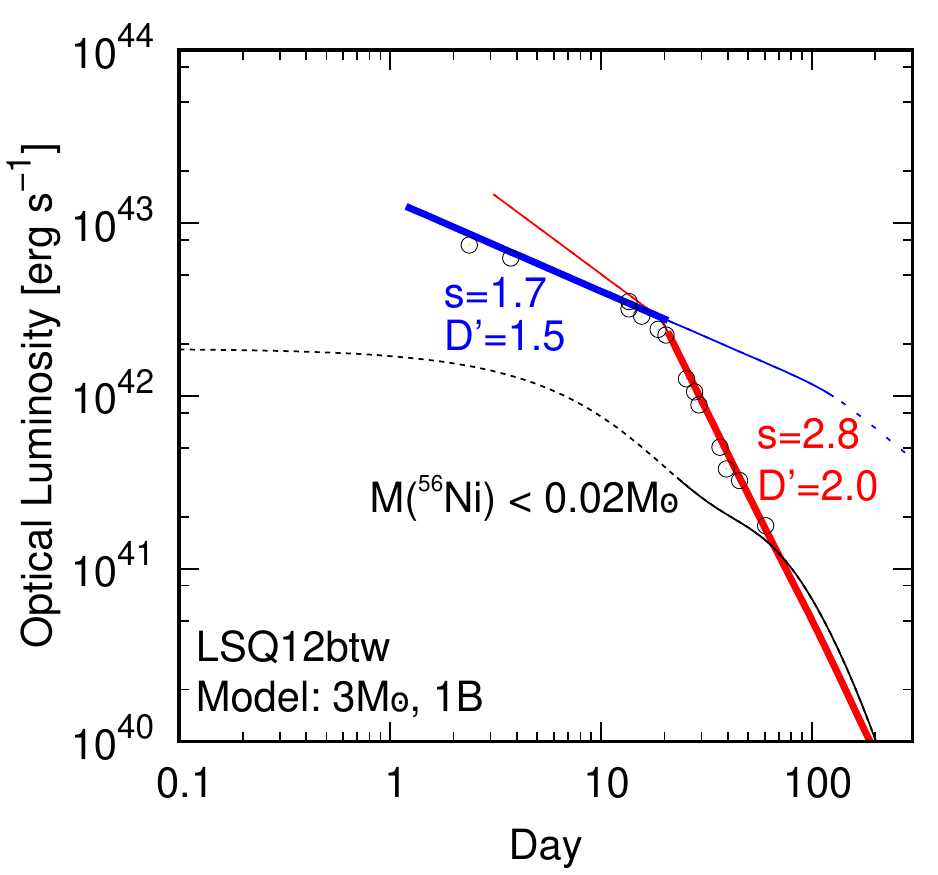}
\caption{The light curve models for individual SNe Ibn. The name of SNe and the model ejecta properties, ejecta mass ($M_\odot$) and kinetic energy (Bethe for $10^{51}$ erg), are shown by a label in each panel. The models are shown by red lines, with another component for the inner CSM shown by blue lines if such a component is required. The CSM properties ($s$ for the density slope and $D'$ for the density scale) are shown in the labels. The {\em maximally-allowed} $^{56}$Ni/Co contribution is described by the black lines, where the phase after the expected diffusion time scale for this contribution is shown by the solid curves. For SN 2006jc, two cases are shown with different choice of the peak/explosion date.
}
\label{fig:individual}
\end{figure*}

The argument in the previous section shows that the model presented here can explain the general behavior of the SN Ibn LCs, and the required model parameters are largely constrained even without studying the LCs of individual SNe Ibn; the LCs as a result of the interaction between the high-density CSM with steep density gradient ($s \sim 2.5 - 3$ and $D' \sim 0.5-5$) and typical (or massive) SESN ejecta ($M_{\rm ej} \sim$ a few $M_\odot$ up to $\sim 5 M_\odot$ and $E_{\rm K} \sim 10^{51}$ erg) reproduce the general properties of SN Ibn LCs. To further constrain the nature of the CSM around SNe Ibn, we construct model LCs for individual SNe Ibn.

As clarified in the previous section, the LC properties are sensitive to the CSM properties but not to the ejecta properties. Adopting a particular set of the ejecta properties in the following is thus only for a demonstration purpose. The examples of the fits to the individual SN Ibn LCs are shown in Fig. \ref{fig:individual}. 
%For each object, we first adopt, as a trial, $M_{\rm ej} = 2 M_\odot$ and $E_{\rm K} = 10^{51}$ erg. The timing of the `kink' in the LC depends on the combination of $M_{\rm ej}$ and $E_{\rm K}$; generally, it takes place earlier for larger $E_{\rm K}/M_{\rm ej}$. When we need to increase the ratio, we increase $E_{\rm K}$, while in the opposite case we increase $M_{\rm ej}$. In case we further need to decrease $E_{\rm K} / M_{\rm ej}$ even if we adopt $M_{\rm ej} = 3 M_\odot$, then we start decreasing $E_{\rm K}$. In the opposite case in which we increase $E_{\rm K}$ while fixing $M_{\rm ej} = 2 M_\odot$, sometimes we encounter the situation where the reverse shock reaches to the inner flat ejecta and thus our model becomes no more applicable. In such a situation, we increase both $E_{\rm K}$ and $M_{\rm ej}$ with a constraint that $M_{\rm ej}$ does not exceed $3 M_\odot$. While the above procedures are adopted to find the model for which $M_{\rm ej}$ falls into the range between $2$ and $3 M_\odot$, it indeed does not guarantee that we have a solution with $M_{\rm ej} = 2 - 3 M_\odot$; we however find reasonable models 
In the examples shown here, we set the ejecta mass either as $2$ or $3 M_\odot$. Then the kinetic energy is varied to match to the early-phase luminosity and the timing of the transition. The CSM properties are largely determined by the luminosity and the slope in the late rapid-decay phase, and they are insensitive to the ejecta properties.

For four out of the seven objects (noting that two cases are shown for SN 2006jc with different choice on the explosion date), only a single CSM component is required. For the remaining three objects, the early-phase evolution is indeed flatter than the model prediction, or is even increasing. This behavior is not explained by the model with a single CSM component, and thus we introduce an additional, flat CSM distribution in the inner part (as shown by the blue lines). For these SNe, it is seen that the timing when the corresponding CSM changes (i.e., the intersection between the red and blue lines) is roughly coincidence with the kink predicted in the outer CSM due to the transition in the shock properties; this is indeed a natural situation as once the shock reaches to the outer steep CSM, the density quickly decreases and thus the shock property changes. 

\section{Nature of SNe Ibn}

\subsection{Circumstellar Environments and Mass-loss History} 

\begin{figure}[t]
\centering
\includegraphics[width=\columnwidth]{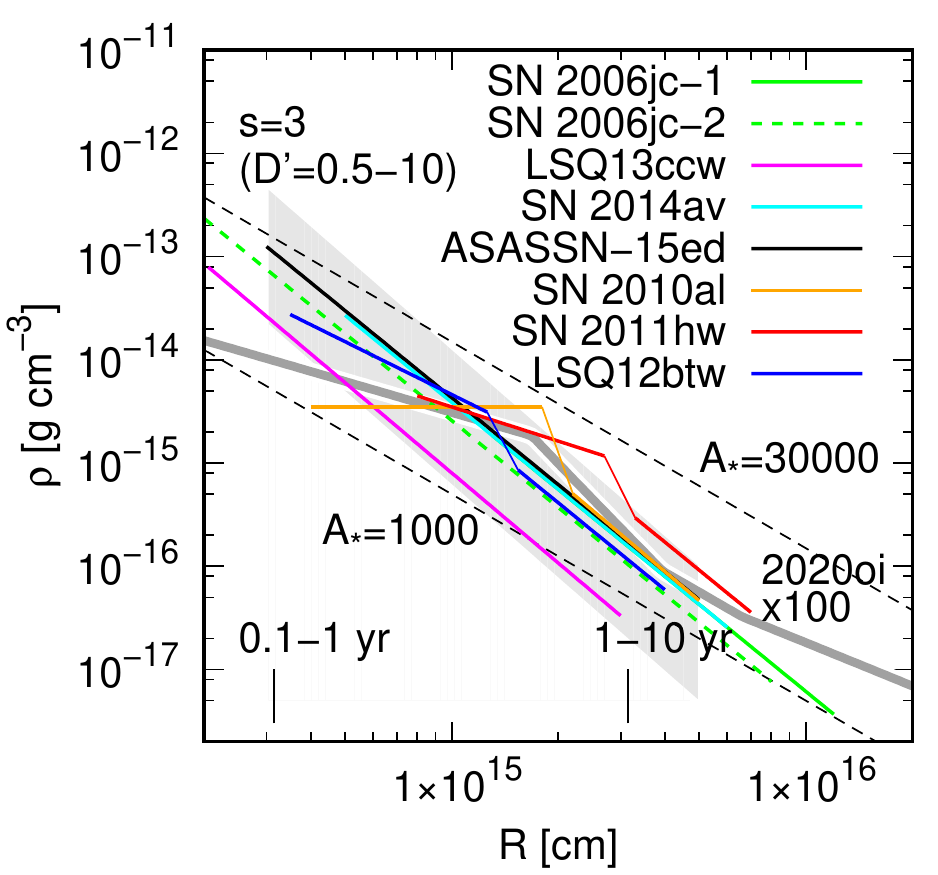}
\caption{The CSM distribution adopted in the LC models shown in Fig. \ref{fig:individual}. For comparison, the CSM density distribution created by the steady-state mass loss is shown for two cases (dashed lines), as described by $\rho_{\rm CSM} = 5 \times 10^{11} A_{*} r^{-2}$ g cm$^{-1}$; $A_{*} = 1$ corresponds to a typical WR wind (i.e., $\dot M = 10^{-5} M_\odot$ yr$^{-1}$ and $v_{\rm w} = 1,000$ km s$^{-1}$), and $A_{*} \sim 5,000 D'$. Also, the CSM distribution with $s=3$ is shown for $D'=0.5 - 10$ by the shaded area, which covers the CSM distributions derived for SNe Ibn. The CSM distribution derived for a canonical SN Ic 2020oi, with the density multiplied by 100, is shown by the gray line \citep{maeda2021}. The corresponding look-back time toward the core collapse is indicated at the bottom for $v_{\rm w} = 100 - 1,000$ km s$^{-1}$. 
}
\label{fig:csm}
\end{figure}

Figure \ref{fig:csm} shows the CSM distribution for a sample of SNe Ibn adopted to explain their LC behaviors. The CSM distribution constrained here is largely in the range from $\sim 3 \times 10^{14}$ cm to $\sim 3 \times 10^{15}$ cm depending on the object. The inner and outer radii correspond to the look-back time of $\sim 0.1 - 1$ yr (for $v_{\rm w} = 100 - 1,000$ km s$^{-1}$) and $\sim 1 - 10$ yr in the mass-loss history, respectively. The CSM distribution is steeper than the steady-state mass loss, following $s \sim 3$ for all SNe Ibn examined in the present work. If the mass-loss velocity would not change much, the steep CSM distribution indicates that the mass-loss rate has been increasing toward the core collapse, roughly following $\dot M \propto (t_0-t)^{-1}$ where $t_0 - t$ is the look back time toward the core collapse. 

Further, the derived CSM density is very high, in the range of $D' \sim 0.5 - 5$. At $\sim 5 \times 10^{14}$ cm, this corresponds to $D \sim (5 - 50) \times 10^{-15} $ g cm$^{-1}$ or $A_{*} \sim 2,500 - 25,000$; $\dot M \sim 0.025 - 0.25 M_\odot$ yr$^{-1}$ if $v_{\rm w} = 1,000$ km s$^{-1}$, or $\dot M \sim 0.0025 - 0.025 M_\odot$ yr$^{-1}$ if $v_{\rm w} = 100$ km s$^{-1}$. We note this is roughly consistent with the estimate using the diffusion time scale as compared to the peak date by \citet{moriya2016}; this constraint is also included in our model as well, and the present work shows that this CSM density is also consistent with the overall LC behaviors. The mass-loss rate derived at $\sim 5 \times 10^{14}$ cm is roughly within the range found for a sample of SNe IIn \citep[$0.001 - 0.1 M_\odot$ yr$^{-1}$ for $v_{\rm w} = 100$ km s$^{-1}$][]{moriya2014a,moriya2014b} but on the highest side. Note however that the evolution of the mass loss is very different; a typical value for SNe IIn is $s \sim 2$ being consistent with the steady-state mass loss \citep{moriya2014a}, while it is $s \sim 3$ for SNe Ibn (this work). 

For three objects out of the seven SNe Ibn, there is a hint of the inner flat CSM component, where the transition radius from the outer steep CSM component is $\sim (1-3) \times 10^{15}$ cm. The behavior is similar to that found for SN Ic 2020oi, albeit the overall flatter and much lower density (by two orders of magnitudes) for SN 2020oi \citep{maeda2021}. The similarity in the characteristic time scale might indicate that they share key physical mechanism as a driver for the intensive mass loss, while the differences would indicate that there is important difference between the progenitors of SNe Ibn and canonical SESNe that leads to the rapidly increasing and very energetic mass-loss driving mechanism for SNe Ibn. 

The CSM mass swept up in 50-200 days after the explosion is typically $\sim 0.2 M_\odot$, covering $\sim 0.05 - 0.4 M_\odot$. Given the accelerated mass-loss activity toward the core collapse in the final $1 - 10$ yrs derived in these models, the estimate here would not be affected even if the longer LC evolution would be observed (unlike the case for SNe IIn, for which the swept-up mass typically keeps increasing as time goes by). The mass lost in this final phase is thus below a typical He layer mass \citep[$\sim 2 M_\odot$, but depending on the rotation, convection, and binarity;][]{nomoto1988,woosley1995,langer2012,laplace2021}. This provides a consistent picture for a WR progenitor for SNe Ibn. This further suggests that the final activity is not a major mechanism for the H and He envelope stripping \citep{fang2019}.

%2006jc... 0.2Msun (case 1) or 0.12 Msun (case 2) at 200 day
%lcq13ccw... 0.05 Msun at 50 day
%sn2014aw... 0.2 Msun
%asassn15ed... 0.2 Msun ay 80 day
%sn2010al... 0.35Msun at 80 day
%sn2011hw... 0.35Msun at 70 adays
%LSQ12btw... 0.12 Msun at 50 days

\subsection{Implications for the Progenitors}

With the properties of the ejecta and CSM quantitatively constrained through the LC models, in this section we discuss possible implications for the nature of the progenitors of SNe Ibn. Before going this far, we also emphasize another important differenece between SNe Ibn and canonical SESNe; it is the mass of $^{56}$Ni ejected at the explosion. 

Previous works have demonstrated that $M$($^{56}$Ni) is generally smaller for SNe Ibn than canonical SESNe \citep{moriya2016,perley2021}. We further quantify this argument. A strong constraint on $M$($^{56}$Ni) can be obtained by the requirement that the bolometric luminosity in the rapidly decaying tail phase should not exceed the $^{56}$Ni/Co contribution (Fig. \ref{fig:individual}). One issue is a possible dust formation that leads to underestimate of the bolometric luminosity if the NIR/IR contribution is not taken into account; we however note that this effect is included for objects showing clear dust-formation signatures, or this effect is considered not to be important in the objects and phases considered in the present works (see Section 2). 

\begin{figure}[t]
\centering
\includegraphics[width=\columnwidth]{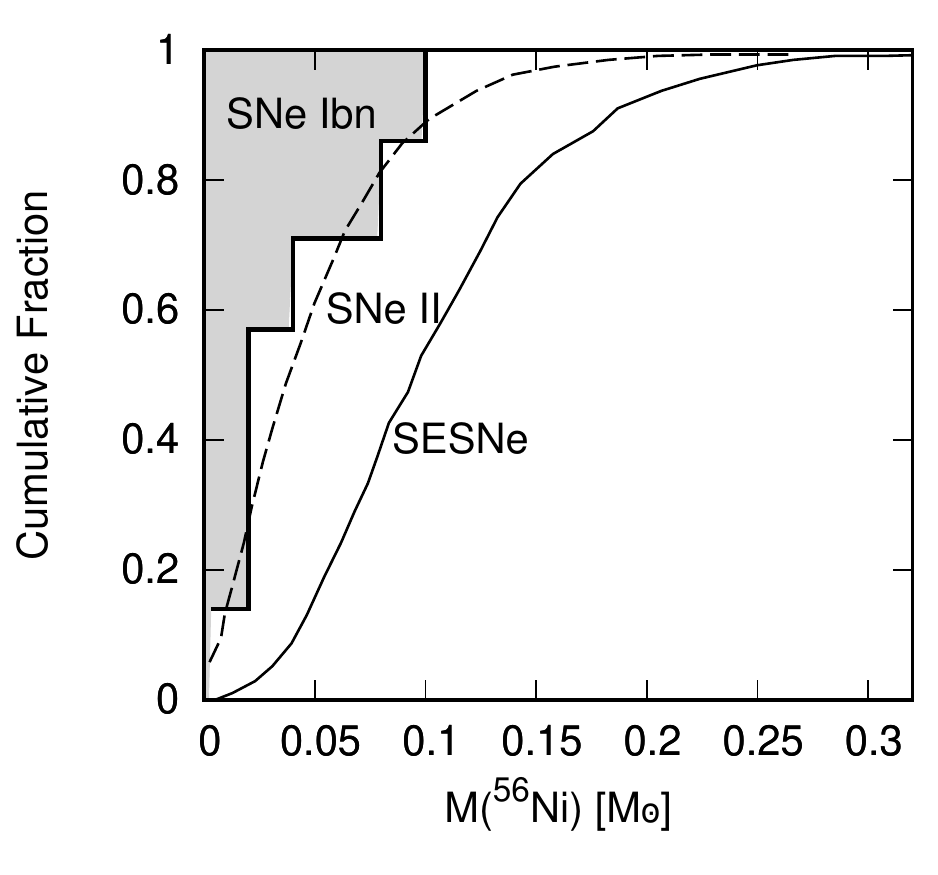}
\caption{Cumulative distribution of $M$($^{56}$Ni). For SNe Ibn, only upper limits are placed for $M$($^{56}$Ni), and thus the gray-shaded area provides an allowed region. The distributions for SNe II (excluding SNe IIn) and SESNe are taken from \citet{meza2020} and \citet{ouchi2021}. Note that the distribution for SESNe is probably biased toward a larger amount of $^{56}$Ni due to a selection effect \citep{ouchi2021} (see the main text). 
}
\label{fig:ni}
\end{figure}

Figure \ref{fig:ni} shows the cumulative distribution of $M$($^{56}$Ni). For SNe Ibn, $M$($^{56}$Ni) is given as an upper limit for all the objects, and the gray-shaded area is the allowed region. The figure also shows the cumulative $M$($^{56}$Ni) distribution for (observed) SESN and SN II samples \citep{meza2020,ouchi2021}. We note that the distributions for SESNe may not represent unbiased, volume-limited samples. While there could be an intrinsic difference in the $^{56}$Ni production between SESNe and SNe II \citep{anderson2019,meza2020}, at least a part of the difference is likely attributed to the selection bias effect; the SN II sample is likely volume-limited so that the observed $^{56}$Ni distribution traces the intrinsic property, while $M$($^{56}$Ni) of SESNe is systematically overestimated and it is possible that the `intrinsic' $^{56}$Ni distribution of SESNe may not be very different from that of SNe II \citep{ouchi2021}. 

Fig. \ref{fig:ni} shows that $M$($^{56}$Ni) ejected in SNe Ibn is clearly smaller than that of the (observed) SESN sample, confirming the previous claim. We further note that the $M$($^{56}$Ni) distribution for SNe Ibn is very likely even skewed to the lower value than SNe II. While the sample is very small, it is very surprising if all the upper limits placed for SNe Ibn would be close to the real values so that the distribution would be marginally consistent with that for SNe II. Therefore, even if we consider a possible selection bias for SESNe, we conclude that the difference between SNe Ibn and SESNe should be real; SNe Ibn produce a smaller amount of $^{56}$Ni than other SESNe (and even than SNe II). Given that the $^{56}$Ni production is controlled by the progenitor core structure and the explosion mechanism \citep{maeda2009,sukhbold2016,sawada2019,suwa2019}, the finding here indicates that there must be an intrinsic difference in the progenitors of SNe Ibn and other canonical CCSNe (i.e., SESNe and SNe IIP). 

\begin{figure*}[t]
\centering
\includegraphics[width=\columnwidth]{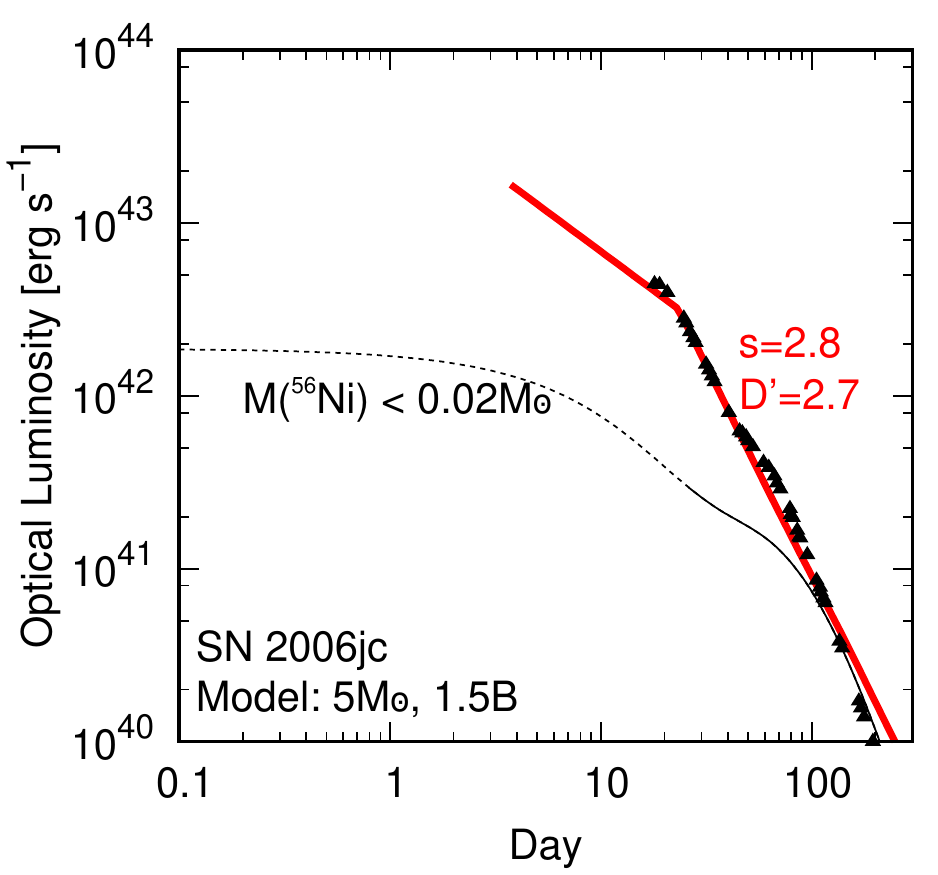}
\includegraphics[width=\columnwidth]{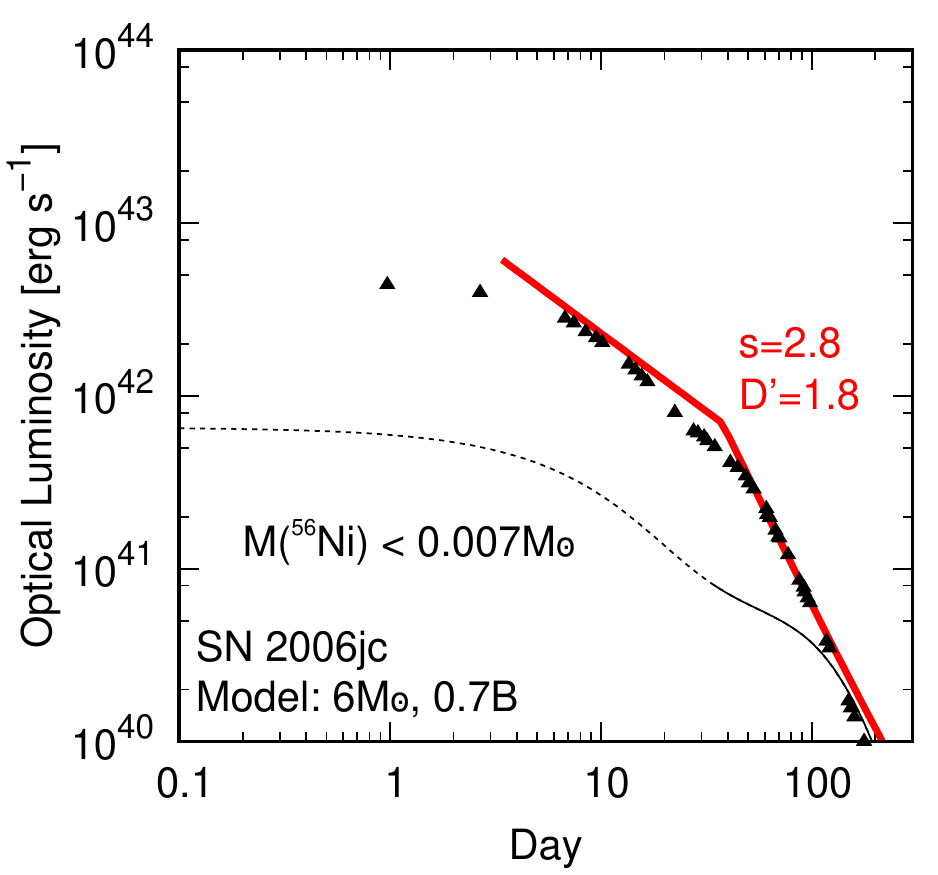}
\caption{The light curve models for SN Ibn 2006jc, adopting more massive ejecta than in the reference models. The two panels are for different choice of the peak date (or explosion date). 
}
\label{fig:massive}
\end{figure*}

The ejecta properties have been constrained to be largely consistent with those for (canonical) SESNe. 
%The less energetic explosion with smaller ejecta mass (i.e., $E_{\rm K} \ll 10^{51} erg$ and $M_{\rm ej} \lsim 1 M_odot$) is not favored for a bulk of SN Ibn population. The other extreme case of highly energetic explosion with massive ejecta ($E_{\rm K} \gg 10^{51}$ erg and $M_{\rm ej} \gsim 5 M_\odot$) does not reproduce typical SN Ibn light curves either. 
As a variant of the SESN ejecta properties, we note that the ejecta as massive as $\sim 6 M_\odot$ is still possible in reproducing the SN Ibn LCs, as long as $E_{\rm K} \sim 10^{51}$ erg (up to a few $\times 10^{51}$ erg), as shown in Figure \ref{fig:massive}. 

In summary, we have the following constraints on the nature of SN Ibn progenitors and explosion mechanism(s): 
\begin{itemize}
\item {\bf Ejecta: } $M_{\rm ej} \sim 2 - 6 M_\odot$ and $E_{\rm K} \sim 10^{51}$ erg (see Section 3.2 for the degeneracy between $M_{\rm ej}$ and $E_{\rm K}$). Namely, the ejecta properties are consistent with those derived for canonical SESNe, but relatively massive ejecta (but with a canonical explosion energy) are also allowed in view of the LC evolution. 
\item {\bf CSM: } Generally steep density gradient ($s \sim 3$), which indicates that the mass-loss rate has increased toward the time of the explosion. The CSM density in the inner region ($\lsim 10^{15}$ cm) is similar to that derived for SNe IIn while the steep distribution for SNe Ibn is very different from SNe IIn. The CSM density is much higher than those derived for SESNe and SNe IIP. In summary, there is no counterpart in known SN populations showing a combination of the very high mass-loss rate and the rapid increase of the mass-loss rate toward the explosion as indicated for SNe Ibn. 
\item {\bf $^{56}$Ni production: } SNe Ibn eject no or little $^{56}$Ni, with the upper limits significantly lower than $M$($^{56}$Ni) in canonical CCSNe (i.e., SNe IIP and SESNe). Given that the production of $^{56}$Ni is linked to the progenitor (either directly or indirectly through the explosion mechanism), there must be an intrinsic difference in the nature of the progenitors between SNe Ibn and canonical CCSNe. 
\end{itemize}

There has been a consensus on the nature of SN IIP progenitors, especially thanks to the number of progenitor detection in pre-SN images \citep{smartt2009,smartt2015}; they are red-supergiants (RSGs) with $M_{\rm ZAMS} \sim 8 - 18 M_\odot$. The case is less certain for (canonical) SESNe, but it has been suggested that a bulk of the SESN population shares the similar ZAMS mass range with SNe IIP \citep{lyman2016} and the difference between SESNe and SNe IIP is largely attributed to the strong binary interaction effect for SESNe without which the progenitor would explode as an SN IIP rather than an SESN \citep{fang2019}. As a variant of SESNe, there is a population of SNe Ic-BL and GRB-SNe; while the nature of their progenitors are even less certain, a popular suggestion is that they are a highly energetic explosion of massive WR stars, probably more massive than canonical SESNe and $M_{\rm ZAMS} \gsim 20 M_\odot$ \citep{iwamoto1998,woosley2006}. 

While the difference in the nature of the CSM between SNe Ibn and SESNe could be related to the details of the binary evolution (e.g., the timing of the strong interaction), the effect of the binary alone would not produce the difference in the $^{56}$Ni production as the Fe core structure is mainly controlled by the ZAMS mass \citep{nomoto1988,woosley1995}. A more straightforward interpretation is that SNe Ibn and SESNe are indeed distinct in their progenitors, probably in the initial mass range. The difference in the nature of the CSM, specifically the rapid increase of the mass-loss rate toward the SN in the final several years as derived for SNe Ibn, may then be viewed as a more extreme case of what has been inferred for SESNe \citep{maeda2021}. Assuming that the ZAMS mass range for (canonical) SESNe is largely overlapping with that for SNe IIP ($\sim 8 - 18 M_\odot$) \citep{lyman2016}, we may consider either the low-mass range ($\sim 8 M_\odot$) or the high-mass range ($\gsim 18 M_\odot$) for SNe Ibn, if they are distinct in the ZAMS mass from SESNe. The low-mass interpretation is not preferred as they will result in the low-mass ejecta and a low explosion energy \citep[e.g., USSNe;][]{tauris2017}, being inconsistent with the constraints derived in the present work. 

The high-mass WR progenitor is indeed one of the popular scenarios so far suggested for SNe Ibn \citep{pastorello2007,tominaga2008}. For example, a star with $M_{\rm ZAMS} \sim 25 M_\odot$ will produce a $8 M_\odot$ He star or $6 M_\odot$ C+O star, which results in $M_{\rm ej} \sim 4.6 - 6.6 M_\odot$ if a neutron star (NS) is left behind. If a black hole (BH) is formed following the SN explosion, the ejecta mass can be even smaller and could be similar to canonical SESNe. In what follows, we argue that this scenario is consistent (or at least not inconsistent) with the constraints derived from the present LC analysis. 
\begin{itemize}
\item The ejecta mass is within the range allowed by the LC analysis.  
\item The binding energy of the He or C+O core in this scenario is $\sim 10^{51}$ erg. Therefore, if the canonical explosion energy ($\sim 10^{51}$ erg as constrained by the LC analysis) is realized in the SN explosion following the NS formation, it will suffer from a substantial amount of fallback to the NS that may (or may not) be converted to a BH. Because of the fallback, no or little $^{56}$Ni will be ejected \citep{woosley1995,maeda2007}. 
\item They produce a large core with a large luminosity during the evolution. While a relation between the nature of the core and the final evolution is yet to be clarified \citep{fuller2017,fuller2018}, the high luminosity (or a high Eddington ratio) might result in a high level of the activity in the final evolution just before the core collapse, thus in the dramatic increase of the mass-loss rate toward the SN. 
\end{itemize}

The massive stars with $M_{\rm ZAMS } \gsim 18 M_\odot$ can, indeed should, become a bare He or C+O star even without a binary interaction thanks to the strong stellar wind \citep{heger2003,langer2012}, and in principle there should be a population of transients that represents explosions of this type of the stars in their final phase. One caveat is that it is also possible that they would directly collapse to a BH with no or weak SN explosion \citep{lovegrove2013,sukhbold2016,basinger2021,neustadt2021}. However, from the Galactic chemical evolution view point, it has been argued that a large fraction of them should eject most of the C+O core material \citep{suzuki2018}, which requires an SN explosion with the explosion energy at least an order of the core binding energy, i.e., $\sim 10^{51}$ erg \citep{maeda2007}.

While some slowly-evolving SESNe \citep{valenti2012,taddia2016,barbarino2021}, as well as SNe Ic-BL \citep[][and references therein]{woosley2006}, have been suggested to have the high-mass WR progenitor, there is a clear deficiency of these massive progenitors with $M_{\rm ZAMS} \sim 18 M_\odot$ as compared to the expectation from a standard Initial-Mass Function \citep[e.g.,][]{valenti2012}. The high-mass progenitor scenario for SNe Ibn may thus be a missing piece to fill in the gap between the observed SN population and stellar evolution theory. It may also explain the so-called RSG problem, i.e., a lack of SN IIP progenitor with $M_{\rm ZAMS} \gsim 18 M_\odot$ \citep{smartt2009,smartt2015}. 

In the high-mass progenitor scenario, SNe Ibn may represent a large, or at least non-negligible, fraction of end-products of massive star with $M_{\rm ZAMS} \gsim 18 M_\odot$. They may branch into the different phenomena (SNe Ibn, slowly-evolving SESNe, and SNe Ic-BL) by additional parameters in stellar evolution such as the rotation, which may further be related to the binarity \citep{yoon2005,mandel2016}. We note that there is a hint of intensive mass-loss just before the explosion for at least a fraction of slowly-evolving SESNe and SNe Ic-BL \citep{taddia2016,ho2019}. 
%SNe Ic-BL do not show a hint of very dense and confined CSM (REF); while they may be partly masked by the contribution by $^{56}$Ni power, it is likely that the final evolution may also be affected by the additional function beyond the stellar mass (such as rotation) if they are to be explained by the (unified) high-mass progenitor scenario. 

Recently, a class of `SNe Icn', showing the signature of strong interaction with C/O-rich CSM, has been discovered \citep{fraser2021,gal-yam2021,perley2021}. A massive WR progenitor has been suggested, in which a BH might be formed through the fallback accretion \citep{perley2021}. This is basically the same scenario as suggested here for SNe Ibn (while a NS formation is not rejected in the present work). A possible relation between SNe Ibn and Icn in the nature of the progenitor is highly interesting. For example, in the single massive star evolution scenario for $M_{\rm ZAMS} \gsim 18 M_\odot$, there is a transition mass below/above which the final state is a He/C+O star. SNe Icn may thus be a variant of SNe Ibn in the highest mass range in this scenario. Given the SN Icn rate about a factor of 10 smaller than that for SN Ibn \citep{perley2021}, then the transition mass should be $M_{\rm ZAMS} \sim 40-50 M_\odot$ for the standard Initial Mass Function \citep{kroupe2001} (see also Section 6.2). Alternatively, other factors, e.g., rotation and/or binary, may be at work and then the difference in the ZAMS mass may not be so large. We plan to apply our model to LCs of SNe Icn and investigate possible difference in the natures of the ejecta and CSM between SNe Ibn and Icn, which will be key to constraining the progenitor scenario for SNe Icn. 

A challenge for the massive star explosion scenario has been raised for PS1-12sk, showing no detectable star-formation activity at the explosion site \citep{sanders2013,hosse2019}. Its peak luminosity is roughly $\sim 10^{43}$ erg s$^{-1}$ and the rising time scale is $\gsim 10$ days. Putting these observational features into the present LC model framework, we infer that it is difficult to explain its LC by a combination of low explosion energy and small ejecta mass (e.g., $\sim 10^{50}$ erg and $0.5 M_\odot$). This can be a strong constraint for a white-dwarf (WD) thermonuclear explosion origin for PS1-12sk; a partial explosion of a WD \citep[e.g., `.Ia'][]{bildsten2007} predicts a small ejecta mass and a low explosion energy (as compared to canonical SNe). To satisfy the constraints derived here, it likely requires a whole disruption of a WD (i.e., SNe Ia) under the WD thermonuclear scenario. We note that no strong constraint has been placed on $M$($^{56}$Ni) for PS1-12sk; $\lsim 0.5 M_\odot$ or $\lsim 1.5 M_\odot$ depending on how it is estimated \citep{sanders2013}. This may then be a counterpart of `SNe Ia' and `SNe Ia-CSM' \citep{hamuy2003,dilday2012}, similarly to the other SNe Ibn related to SESNe and SNe IIn.  

\section{Applications to Rapidly Evolving Transients}

\begin{figure*}[t]
\centering
\includegraphics[width=\columnwidth]{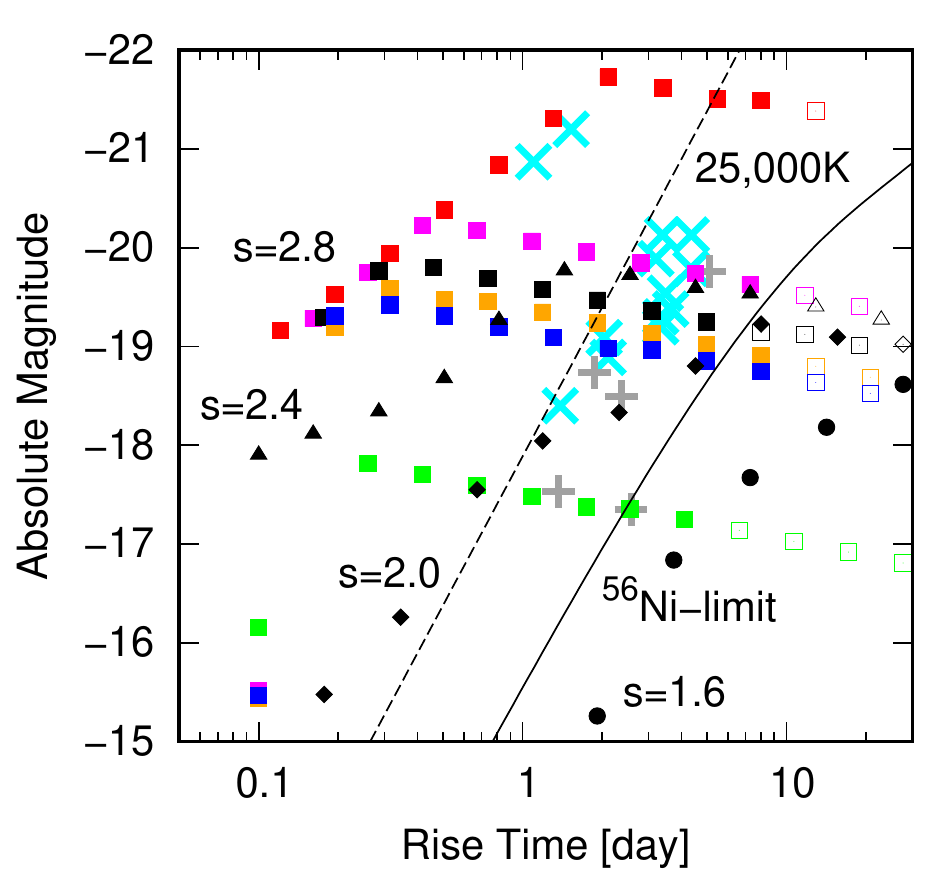}
\includegraphics[width=\columnwidth]{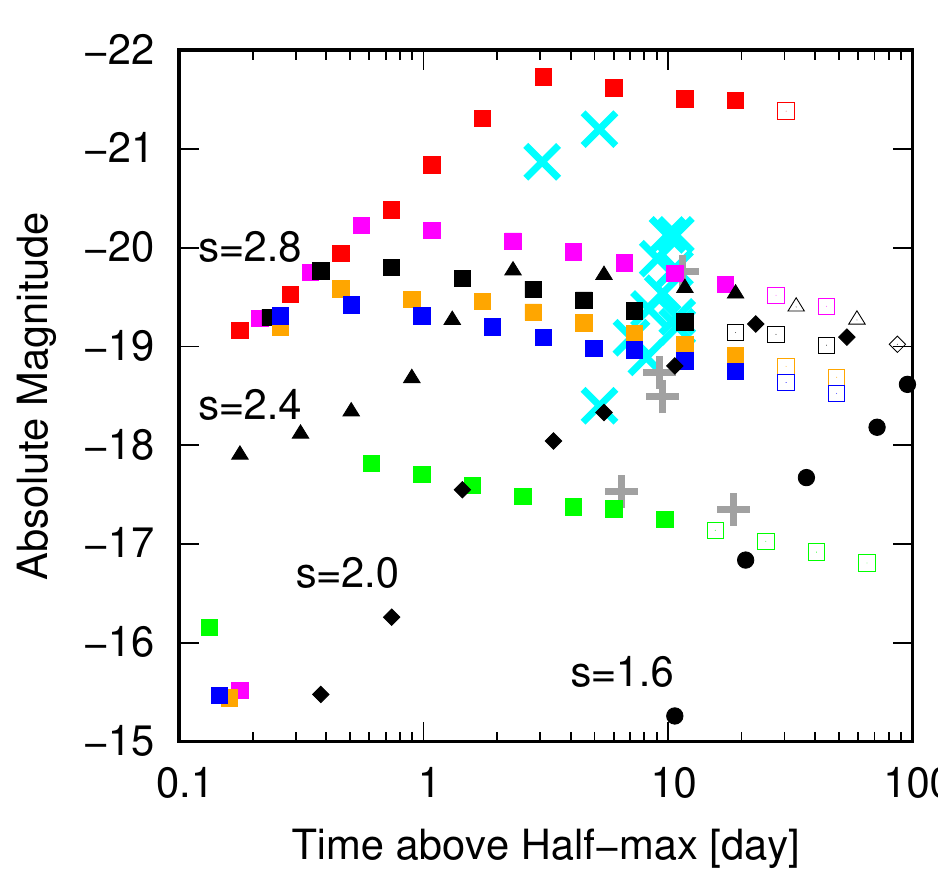}
\caption{The relation between the LC time scale and the peak absolute magnitude. The time scale is shown for the rise time (left) or for the time above half-max as defined by the duration in which the transient is brighter than the half of the peak luminosity (right). The square, black-triangle, black-diamond, and black-circle symbols are for the models (see the main text for details). For comparison, we plot the ZTF samples of `SN Ibn' rapid transient population (including some literature sample) \citep[cyan-crosses:][]{ho2021}, and the HSC sample of rapid transients \citep[gray-crosses:][]{tampo2020}. The dashed line indicates the photospheric temperature of $25,000$ K for the velocity of $15,000$ km s$^{-1}$ typically found in the models. The solid curve defines the critical line above which the required $^{56}$Ni mass exceeds the ejecta mass. 
}
\label{fig:tpeak_mag}
\end{figure*}

The recent progress in high-cadence surveys has been expanding the phase space in the transient observations, leading to discovery of various new classes of objects. One of the most exciting findings is the existence of rapidly-evolving transients with the characteristic time scale of $\lsim 10$ days, which had been missed previously in traditional surveys. A few examples had been discovered by chance \citep{kasliwal2010,ofek2010,poznanski2010} with diverse observational properties including the spectral classification type, indicating multiple populations within the rapidly-evolving transients. 
It has only recently become possible to systematically search for the rapidly-evolving transients by high-cadence surveys \citep{drout2014,tanaka2016,pursiainen2018,tampo2020}. However, these survey samples were not based on the real-time transient identification, and thus spectral classification has been largely unavailable. 

Very recently, \citet{ho2021} presented 42 rapidly-evolving transients from the Zwicky Transient Facility (ZTF) survey \citep{bellm2019}, with about half classified spectroscopically. This allows the first characterization of a uniform sample, and they confirmed the existence of multiple populations, largely divided into three categories; sub-luminous SN IIb/Ib population, luminous SN Ibn population, and fast and blue optical transients (FBOT) \citep[or AT2018cow-like transients;][]{margutti2019,perley2019}. 

Therefore, SNe Ibn form one of the major populations among the rapidly-evolving transients. Indeed, the opposite seems also the case; a large fraction of SNe Ibn belong to the rapid transient population \citep{moriya2016,hosse2017}. It is therefore of great interest to compare a grid of the present models and the observations of the SN Ibn subclass of the rapidly-evolving transients. 

\subsection{Comparison to `SN Ibn' Rapidly-Evolving Transients}

Figure \ref{fig:tpeak_mag} shows the relation between the LC time scales (the rise time and the time above the half-max) and the peak magnitudes. The square symbols are for the models. The models with $s=2.8$ are shown for various combinations of the ejecta properties; $(M_{\rm ej}/M_\odot, E_{\rm K}/10^{51} \ {\rm erg}) = (6, 10)$, $(4, 2)$, $(2, 1)$, $(3, 1)$, $(0.5, 0.1)$, following the same color coordinate as in Fig. \ref{fig:ejecta}. Additionally, models with $s=2.4$, $2$, and $1.6$, all with $M_{\rm ej} = 2 M_\odot$ and $E_{\rm K}/10^{51} \ {\rm erg} = 1$, are shown by the black triangles, diamonds, and circles, respectively. For each model sequence, different points are for different values of $D'$, ranging from $32$ from the right to 0.033 to the left, with a step of a factor of two each. Additionally, the models with $D'=0.003$ are shown, which are mostly overlapping in the left-bottom corner in the panel. For comparison, we plot the ZTF (plus literature) samples of `SN Ibn' rapid transient population \citep[cyan-crosses;][and references therein]{ho2021}, and the Subaru-HSC sample of rapid transients \citep[gray-crosses;][]{tampo2020}. Note that the observational points are either for the observed $g$ or $i$ band, while the model points are for the bolometric LC. 

The characteristic properties in the time scale and the peak magnitude are well explained by the present model sequence, with the parameter set similar (or basically identical within the uncertainty) to those applied to the SN Ibn sample in the previous sections (i.e., $s \sim 2.8$, $D' = 1 - 4$, with SESN or massive SESN ejecta properties). It is not surprising, as the SN Ibn sample examined in the previous section has the peak magnitude ($\sim -19$ mag) and the characteristic time scale ($\lsim 10$ days) largely overlapping with the typical properties of the ZTF SN Ibn rapid population. The HSC sample, which are not spectroscopically classified, seems to be a mixture of the `SN Ibn' and `faint SN IIb' populations, where the former overlaps with the ZTF sample in their LC  properties. 

\begin{figure}[t]
\centering
\includegraphics[width=\columnwidth]{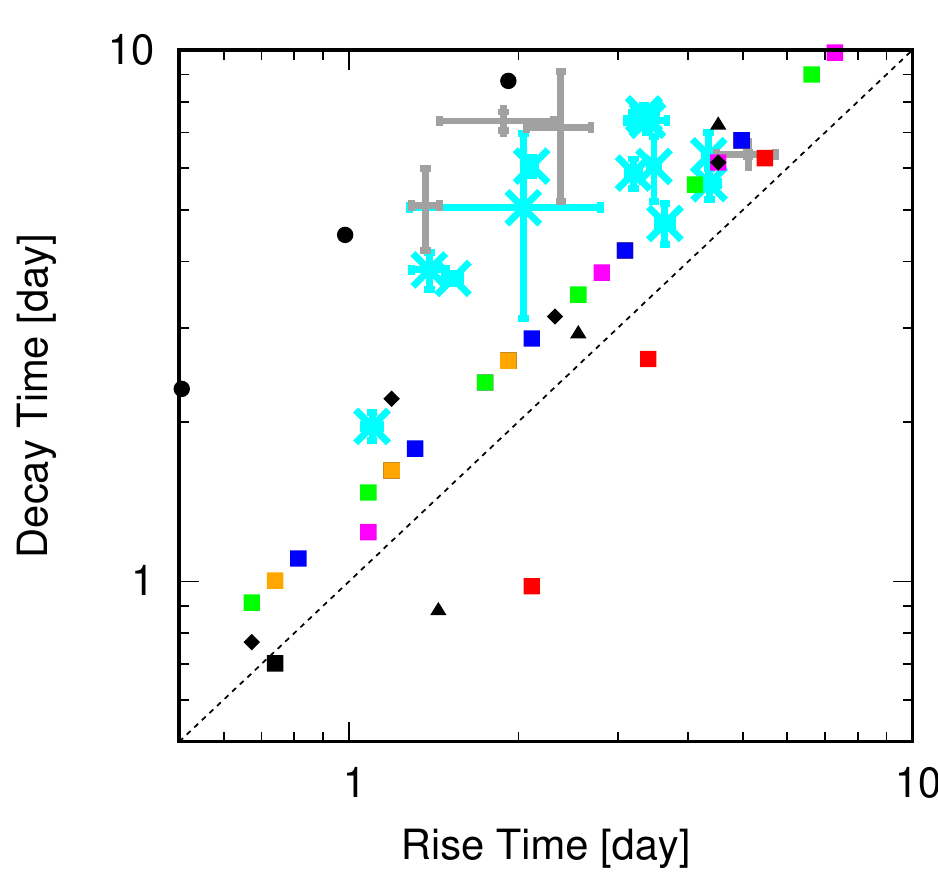}
\caption{The relation between the rise time and the decay time (the duration in which the luminosity decreases by 50\% after the peak). The models and observed samples are the same with those in Fig. \ref{fig:tpeak_mag}. 
}
\label{fig:rise_decay}
\end{figure}

Figure \ref{fig:rise_decay} compares the rise time and decay time. The observed sample of rapidly-evolving transients generally shows that there is a rough correlation between the two, while the decay time scale is longer than the rise time. While our estimate on the rise time in the model is very crude since we have not solved the diffusion process directly, qualitative behavior is reproduced. Interestingly, a ratio of the decay time to the rise time is controlled by the CSM slope ($s$) and is insensitive to the ejecta properties ($M_{\rm ej}$ and $E_{\rm K}$) and the CSM density scale ($D'$). The exception for this behavior is found for an energetic ejcta model (red squares); the model predicts vert rapid decay, since the high shock velocity leads to the transition to the adiabatic shock already at the peak (Section 3.2). The predicted behavior for the energetic model thus does not fit to the observed properties of the rapidly evolving transients. 

As outliers, there are two luminous and rapid SNe Ibn in the ZTF/literature sample ($\sim -21$ mag and $\sim 1$ day for the rise time) shown in Fig. \ref{fig:tpeak_mag} that seem to be distinct from the remaining SNe Ibn. They overlap with the AT2018cow-like (FBOT) transients in the phase space; indeed, one of them is AT2018cow itself, for which \citet{ho2021} gave a label of `SN Ibn?' according the classical spectral classification, but they noted that this is indeed very different from SNe Ibn in many features. The other one is iPTF15ul, but its spectra are heavily contaminated by the host galaxy and thus the SN Ibn classification is very tentative \citep{hosse2017}. Therefore, indeed these two outliers may not be SNe Ibn at all. The origin of AT2018cow has been discussed intensively with various scenarios suggested, but so far no firm conclusion has been reached. Indeed, our SN Ibn models with standard ejecta properties fail to explain the LC characteristics of these FBOTs. It is  interesting that the CSM interaction between the dense CSM ($D' \sim 1$) and the `SN Ic-BL / GRB-SN' ejecta could explain their LC characteristics, i.e., the combination of the high peak luminosity and very short time scale. However, such a scenario has various difficulties in explaining the best-observed FBOT AT2018cow \citep[e.g.,][]{uno2020}, and also unable to explain the relation between the rise and decay time scales (see above). Still, an energetic version of SNe Ibn in the present scenario (i.e., SNe Ic-BL within a confined CSM) may contaminate to these luminous FBOTs, if they would exist in nature. It is thus interesting to further investigate if the properties of the FBOTs are uniform or diverse, where the latter may probe a possible contamination of an energetic version of the interaction-powered SNe Ibn. 

Advantage of performing comparison between the model grid and uniform sample is the possibility to constrain the model parameter space that would in turn provide a hint for the progenitor evolution toward SNe Ibn and a part of the rapid transient population. Fig. \ref{fig:tpeak_mag} shows that the ejecta properties are basically limited to those of typical SESNe and potentially massive SESN ejecta; the ejecta properties corresponding to SNe Ic-BL (or GRB-SNe) and less-massive SESNe (e.g., USSNe) are largely rejected for the SN Ibn rapid population (but see above for the `FBOT' population). There seems no strong bias such that they would be missed if they would exist in nature (e.g., fainter `SNe IIb/Ib' rapid population has been detected). Similarly, the distribution is consistent with the steep CSM distribution ($s > 2$), suggesting the steep CSM as a generic feature of the SN Ibn rapid population. 

Interestingly, the observed sample of SN Ibn rapid population has a limited range in the CSM density ($D' \sim 1-4$) if they are interpreted as the SN-CSM interaction. This has various implications for the origin(s) of SNe Ibn and the related rapid population. If there would be events with very high mass-loss rate exceeding $D' \sim 4$, there seems no particular observational bias with which they are missed from the survey and follow-up observations. Therefore, the upper limit of $D' \sim 4$ is probably real and this must be a strong constraint on any progenitor evolution scenario for SNe Ibn. This may be the maximum level of the pre-SN activity that such a massive WR star can have. An yet another possibility is that this limit could be related to SNe Icn; the mass-loss rate significantly exceeding $D' \sim 4$ will eject all the He envelope, and further mass loss may create the C/O-rich CSM and result in SNe Icn. 

On the other hand, the lower limit, $D' \sim 1$, may represent either an observational bias or real properties of SN Ibn progenitors. If the latter is the case, this makes the SN Ibn progenitor very distinct from canonical SESNe. On the other hand, if the former is the case, the SN Ibn progenitor can be still distinct from SESN progenitors, but the transition from SESNe to SNe Ibn may take place in a continuous manner. We will discuss further detail on the possible observational bias in the next section. 

\subsection{Possible UV Population}

Figure \ref{fig:tpeak_mag} provides an interesting possibility on the possible observational bias to account for a lack of SNe Ibn and/or SN Ibn rapid population with relatively less-dense CSM ($D' \lsim 1$). In the context of the present SN-CSM interaction scenario, a less-dense CSM results in a brighter and shorter event with a smaller photosperic radius at the peak. 
They are thus associated with higher photospheric temperature. Within the SN-CSM interaction scenario, we can estimate the photoshepric temperature ($T_{\rm eff}$) according to the Stefan-Boltzman relation, $L_{\rm bol} = 4 \pi v^2 t^2 \sigma T_{\rm eff}^4$, where $v$ is the velocity at the contact discontinuity and $t$ is the time elapsed from the explosion (noting that this is generally an underestimate for the photosperic radius as the optical depth within the CSM may not be negligible). Adopting a typical velocity of $\sim 15,000$ km s$^{-1}$ for the present models, a line corresponding to $T_{\rm eff} = 25,000 K$ is shown in Fig. \ref{fig:tpeak_mag}. Interestingly, this line seems to define the limit for the SN Ibn population, above which (i.e., at higher temperatures) no SNe Ibn are found in this diagram (note that there are two exceptions belonging to FBOTs, for which the present model would not apply; see above). 

\begin{figure}[t]
\centering
\includegraphics[width=\columnwidth]{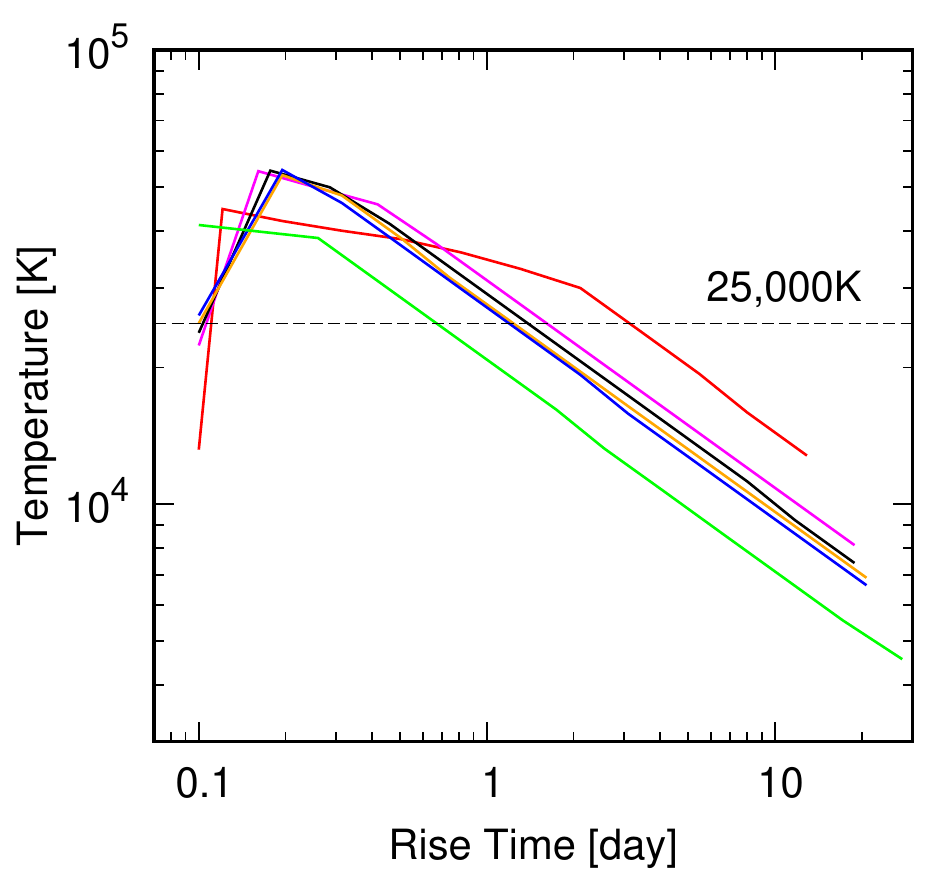}
\caption{The estimated photospheric temperature in the models with various ejecta properties and $D'$, but with $s$ fixed to $2.8$. The optical depth within the CSM is taken into account. The same color coordinate as in Fig. \ref{fig:ejecta} is used to distinguish different ejecta properties. 
}
\label{fig:peakT}
\end{figure}

This raises an interesting possibility. Even if there could be SNe Ibn with relatively less-dense CSM ($D' \lsim 1$) than applied to the present samples, they may emit mostly in the UV and thus such a population may be underrepresented in the present sample constructed based on optical surveys. Figure \ref{fig:peakT} further presents model details, where $T_{\rm eff}$ at the peak is estimated based on the photospheric radius calculated in each model. For a range of the ejecta properties except for low/high-energy models, the relation between the rise time and the temperature at the peak turns out to be universal. The temperature generally increases toward the event with shorter rise time, and it is $\gsim 25,000 K$ for objects with the rise time less than 1 or 2 days. 

Assuming that a large fraction of the radiation is emitted in the UV at such a high temperature, this hypothesized population may be characterized by the peak (UV) luminosity of $\gsim 10^{43}$ erg s$^{-1}$ and the time scale of at most a few days (or even less than 1 day for most of them). Assuming that a typical bolometric magnitude of $\sim 10^{43}$ erg is mostly emitted in the NUV ($\sim 2500$\AA), the detectable horizon is estimated to be $\sim 2$ Gpc for the limiting magnitude of 23 mag (AB). This is $\sim 400$ Mpc even for the 20 mag (AB). Future UV facilities like ULTRASAT \citep{sagiv2014}, with daily all-sky survey down to $\sim 23$ mag in NUV\footnote{https://www.weizmann.ac.il/ultrasat/}, will thus provide a complete census of the possible UV rapid-transient population, and will definitely answer whether the final mass-loss rate for SNe Ibn is indeed continuously distributed toward the lower rate than those so far discovered in the optical. Shallower surveys (e.g., down to 20 mag) will already reach to $z \sim 0.1$ and will start discovering these UV population if they would exist in nature. 

Indeed, if the scenario suggested here is correct, we would expect that there could be even a larger number of `UV' rapid Ibn population than so far discovered in the optical. The volumetric rate of the rapid population has been estimated to be $\sim 5$\% of CCSNe \citep{drout2014}. They are dominated by the SN IIb/Ib population, and the contribution of the SN Ibn population is $\sim 10$\% \citep{ho2021}. Therefore, the rate of the rapid SN Ibn population is $\sim 0.5$\% of the CCSN rate. The rates of SNe Ibn classified into rapid and non-rapid populations seem comparable \citep[see, e.g., Table 1 of ][]{moriya2016}, thus we estimate that the SN Ibn volumetric rate observed so far is $\sim 1$\% of CCSNe. This is far below the fraction of massive stars with $M_{\rm ZAMS} \gsim 18 M_\odot$ to those with $M_{\rm ZAMS} \gsim 8 M_\odot$, which is $\sim 20$\% for the standard Initial Mass Function \citep{kroupe2001}, by an order of magnitudes. While it is possible that not all massive stars with $\gsim 18 M_\odot$ explode as SNe Ibn (or even as any classes of SNe), it is very likely that a number of events are indeed left undetected so far in the optical, if a majority of SNe Ibn are the end-products of massive stars with $M_{\rm ZAMS} \gsim 18 M_\odot$. 

\section{Summary}\label{sec:summary}

In this paper, we present a light curve model for SNe powered by interaction between the ejecta and CSM, taking into account the conversion of the dissipated kinetic power to optical (and UV/NIR) thermal emission. The model is applied to a sample of SNe Ibn and `SN Ibn' rapidly-evolving transients. For SNe Ibn, our findings are summarized as follows: 
\begin{itemize}
    \item The characteristic post-peak behavior commonly seen in the SN Ibn LCs, where a slow decay is followed by a rapid decay, is naturally explained by the transition of the forward-shock property from cooling/radiative to adiabatic regime without introducing a change in the CSM density distribution. 
    \item The steep decay in the light curves of SNe Ibn indicates a steep CSM density gradient ($\rho_{\rm CSM} \propto r^{-3}$). 
    \item The CSM properties are constrained as follows; $\rho _{\rm CSM} \sim (0.5-5) \times 10^{-14}$ g cm$^{-1}$ at $\sim 5 \times 10^{14}$ cm ($D' = 0.5 - 5$); $\dot M \sim 0.025 - 0.25 M_\odot$ yr$^{-1}$ if $v_{\rm w} = 1,000$ km s$^{-1}$, or $\dot M \sim 0.0025 - 0.025 M_\odot$ yr$^{-1}$ if $v_{\rm w} = 100$ km s$^{-1}$. 
    \item The dense and steep CSM structure naturally leads to the transition from the cooling to adiabatic regime in the forward shock properties as mentioned above. 
    \item The ejecta properties are constrained as follows; $M_{\rm ej} \sim 2 - 6 M_\odot$ and $E_{\rm K} \sim 10^{51}$ erg; the properties are consistent either with those of canonical SESNe or those with more massive ejecta (but with the canonical explosion energy of $\sim 10^{51}$ erg). 
    \item SNe Ibn produce no or little $^{56}$Ni; the mass of $^{56}$Ni is significantly lower than SESNe or even than SNe IIP. 
\end{itemize}

These properties are different from those of canonical SESNe and SNe IIP, indicating that SNe Ibn progenitor is intrinsically different and likely originated in the initial mass range different from these SNe. Assuming that SNe IIP and (most of) SESNe are the end-products of massive stars with $M_{\rm ZAMS} \lsim 18 M_\odot$, the finding reinforces the idea that SNe Ibn are an explosion of a massive WR star with $M_{\rm ZAMS} \gsim 18 M_{\odot}$. We have shown that the expected properties are consistent (or at least not inconsistent) with the properties of the ejecta and CSM derived in the present work. 

The inferred mass-loss properties in the final phase of the progenitor evolution are very unique. The mass-loss rate increases toward the explosion and reaches to the level similar to those inferred for SNe IIn. The characteristic timescale of the pre-SN activity has a similarity to that inferred for SN Ic 2020oi, but the mass-loss rate is much larger for SNe Ibn. The similarity and the difference suggest that the key mechanism that drives the final mass loss may be common, but the magnitude of the activity may be dependent on the core mass. 

We have applied the present model to a sample of the `SNe Ibn' rapidly-evolving transient population. As the sample has been constructed through a systematic survey, the comparison allows to place constraints on the properties of the ejecta and CSM.
\begin{itemize}
    \item The general properties of the SN Ibn rapid population are explained by the present model. 
    \item The constraints on the ejecta properties are essentially the same as derived for a sample of SNe Ibn; either canonical SESN or massive (but non-energetic) ejecta. 
    \item The range of the (final) mass-loss rate is also consistent with that derived for the literature sample of SNe Ibn. 
    \item The upper limit in the mass-loss rate ($D' \sim 4$) should be real; under the massive WR progenitor scenario, this is viewed as the maximum level of the pre-SN activity that such a massive WR star leading to SNe Ibn can have. Indeed, the mass-loss rate significantly exceeding this will eject all the He envelope, and may turn into SNe Icn. 
    \item On the other hand, the lower limit of the mass-loss rate ($D' \sim 1$) may be due to a selection bias. Even if such a progenitor with the mass-loss rate lower than that derived for the `optical' SNe Ibn (and related rapid transients) would exist, they will become UV-bright rapidly-evolving transients with the characteristic luminosity of $\sim 10^{43}$ erg s$^{-1}$ and time scale of about a day. 
    \item We suggest that future {\rm UV} missions such as {\rm ULTRASAT} will potentially discover a number of the currently-missing UV-bright rapid transients. 
\end{itemize}

\acknowledgments

%The authors thank the referee for insightful and constructive comments. 
K.M. acknowledges support from the Japan Society for the Promotion of Science (JSPS) KAKENHI grant JP18H05223 and JP20H04737. K. M. and T. J. M. acknowledge support from the JSPS KAKENHI grant  JP20H00174. The authors thank the Yukawa Institute for Theoretical Physics at Kyoto University; discussion during the YITP workshop YITP-T-21-05 on `Extreme Outflows in Astrophysical Transients' was useful for this work.


%\vspace{5mm}
%\facilities{}


%\software{}



\bibliography{snibn_lc}{}
\bibliographystyle{aasjournal}



\end{document}